 \documentclass[aps,pre,twoside,twocolumn,showpacs]{revtex4}

 \usepackage{amsmath,amssymb,bm,euscript,mathrsfs}
 \usepackage{graphicx,boxedminipage}
 \usepackage[bf,small]{caption2}
 \usepackage{floatflt}
 \usepackage[hypertex]{hyperref}
 \usepackage{xcolor,fancybox}

\def\prep{Phys. Rep.\ }

\def\WRM{Waves Random Media\ }

\newcommand{\bra}[1]{\ensuremath{\bm{\langle}#1\bm{|}}}
\newcommand{\ket}[1]{\ensuremath{\bm{|}#1\bm{\rangle}}}

\newcommand{\av}[1]{\ensuremath{\langle#1\rangle}}
\newcommand{\Av}[1]{\ensuremath{\big<#1\big>}}
\newcommand{\AV}[1]{\ensuremath{\Big<#1\Big>}}

\begin{document}

\title{SPECTRAL PROPERTIES OF CYLINDRICAL QUASI-OPTICAL CAVITY RESONATOR
       WITH RANDOM-INHOMOGENEOUS SIDE BOUNDARY:\\
       CORRELATION BETWEEN DEPHASING AND DISSIPATION}

 \author{E.\,M. Ganapolskii}
 \author{Yu.\,V. Tarasov}\email{yutarasov@ire.kharkov.ua}
 \author{L.\,D. Shostenko}

 \affiliation{Institute for Radiophysics and Electronics,
              National Academy of Sciences of Ukraine,
              12 Proskura Street, Kharkov 61085, Ukraine}
\begin{abstract}
 A rigorous solution for the spectrum of a quasioptical cylindrical cavity resonator with a randomly rough side boundary has been obtained. To accomplish this task, we have developed a method for the separation of  variables in a wave equation, which enables one, in principle, to rigorously examine any limiting case --- from negligibly weak to arbitrarily strong disorder at the resonator boundary. It is shown that the effect of disorder-induced scattering can be properly described in terms of two geometric potentials, specifically, the ``amplitude'' and the ``gradient'' potentials, which appear in wave equations in the course of conformal smoothing of the resonator boundaries. The scattering resulting from the gradient potential appears to be dominant, and its impact on the whole spectrum is governed by the unique sharpness parameter~$\Xi$, the mean tangent of the asperity slope. As opposed to the resonator with bulk disorder, the distribution of nearest-neighbor spacings (NNS) in the rough-resonator spectrum acquires Wigner-like features only when the governing wave operator loses its unitarity, i.\,e., with the availability in the system of either openness or dissipation channels. It is shown that the reason for this is that the spectral line broadening related to the oscillatory mode scattering due to random inhomogeneities is proportional to the dissipation rate. Our numeric experiments suggest that in the absence of dissipation loss the randomly rough resonator spectrum is always regular, whatever the degree of roughness. Yet, the spectrum structure is quite different in the domains of small and large values of the parameter $\Xi$. For the dissipation-free resonator, the NNS distribution changes its form with growing the asperity sharpness from Poissonian-like distribution in the limit of $\Xi\ll 1$ to the bell-shaped distribution in the domain where $\Xi\gg 1$.
\end{abstract}
 \date{\today}
 \maketitle

\section{Introduction}

A great deal of attention is currently being given to studying the spectral properties of microwave systems in which the propagation of waves is highly, or even totally, restricted in all three dimensions. Systems of this sort, widely referred to as zero-dimensional systems, are commonly used in optic, acoustic, and laser technologies as resonators, both of cavity and of open type.

Spectral properties of resonators whose spatial dimension substantially exceed an operational wavelength (the so-called
quasioptical resonators) have for a long time been thoroughly investigated (see, e.\,g., Refs.~\cite{bib:Katsenelenb66,bib:Vainstein88}). However, this applies mainly to resonators of relatively simple form, which allow for exact integrability of corresponding wave problems. Things are getting worse with resonators whose geometrical features do not permit the solution of wave equations using conventional variable separation methods. Such resonators, provided they have no internal inhomogeneities, are, by analogy with classical dynamic systems, attributed to the class of nonintegrable systems of the billiard type.

In studies of billiard dynamical systems, the most commonly used are the methods of classical chaos theory \cite{bib:Zaslavskii84}, which rely substantially on the theory of random matrices \cite{bib:Beenakker97,bib:Guhr98}. Using these methods, spectral properties of different billiard systems with unstable classical dynamics have been profoundly studied, in particular, the well-known Sinai and Bunimovich billiards \cite{bib:Sinai70,bib:Bunimovich79}.
Systems of similar type, with regular and mainly smooth boundaries, were examined experimentally using different microwave methods \cite{bib:Stokmann04}. At the same time, spectral properties of resonators whose boundaries possess essentially random properties have thus far received insufficient attention. The adequate theoretic analysis of these
properties is difficult to perform, mainly, due to the widespread belief that in systems with broken spatial symmetries the variables do not separate, at least with a traditional algebraic approach. This makes it necessary to employ rather approximate, and even phenomenological, methods for spectral analysis of such systems, whose application is far from being adequately substantiated.

In our recent papers, Refs.~\cite{bib:GanErTar06-1,bib:GanErTar06-2,bib:GanErTar07}, using the method developed previously in Ref.~\cite{bib:Tar00} we have shown that one can separate variables in a wave equation and thus perform
a comprehensive spectral analysis, for any restricted wave system, however disordered, lossy, or lossless it may be.
The only requirement for the system in question is that its difference from some deliberately integrable reference system might be described in terms of certain effective potentials. In this case, the variables can always be separated, although not necessarily by the algebraic approach but rather with some operator technique. Nevertheless, this enables one to scrutinize any limiting case, from negligibly weak to arbitrary strong scattering of oscillatory modes due to inhomogeneities in the resonator.

In the above-mentioned papers it was found that the scattering produced by static inhomogeneities smoothly distributed
in the resonator bulk, though Rayleigh in nature, results in additional spectral line broadening, which is of essentially \emph{nondissipative} origin and is additive to the broadening caused by dissipation processes. This kind of broadening is explained by the resonance mode decoherence resulting from coexistence in such systems of two scattering channels, specifically, the intramode (coherent) and the intermode (incoherent) channel. The impact of infill inhomogeneities on the separate resonance lines is of a highly selective character. The most widened and shifted along the frequency axis appear to be those lines in whose vicinity there is the largest number of other resonances.
Meanwhile, the lines positioned relatively sparsely prove to be weakly subjected to inhomogeneities and retain their high quality factors.

Such a specific behavior of spectral lines was interpreted in Refs.~\cite{bib:GanErTar06-1,bib:GanErTar06-2,bib:GanErTar07} as a kind of interaction between resonances. The interaction results in effective rarefication of the initially dense spectrum of the random inhomogeneous resonator, in
the sense that as the disorder increases the number of high-$Q$-factor lines gets progressively lower. The quality factors of the remaining lines are considerably reduced, and some of the lines even become superposed.We have exploited such an effective rarefication of cavity resonator spectrum, which results from randomization of its infill, in simulations of a random laser with single-frequency generation using the quasioptical millimeter wave band resonator~\cite{bib:GanErTar06}.

In this paper, the investigation is undertaken of the cylindrical quasioptical cavity resonator which is randomized
not through inhomogeneous filling but rather through random roughening of its side boundary. Resonators of such a type are physically equivalent to the two-dimensional billiard system conventionally called the Shepelyansky billiard. It was already examined in Refs.~\cite{bib:FrahmShepel97,bib:FrahmShepelyansky97,bib:Sirko2000,bib:Hlushchuk01} in an effort to elucidate the possibility to observe the effects of dynamic localization and quantum ergodicity. Apart from this, Refs.~\cite{bib:Mirlin2000,bib:BlantMirlMuz2001} also dealt with resonance systems having randomly rough boundaries,
specifically, with solid-state quantum dots. To study the current-carrier spectrum, the authors of Refs.~\cite{bib:Mirlin2000,bib:BlantMirlMuz2001} have applied the ballistic $\sigma$ model intended to describe not only universal but also system-specific properties of quantum resonance systems. Yet, carrier scattering due to boundary
roughness was taken into account in these papers phenomenologically, using the concept of the specular reflection factor~\cite{bib:Fuchs38}. This made it impossible for the authors to allow for the diffraction effects produced in the course of quantum wave scattering and, as a consequence, to carry out in-depth spectral analysis for the system where the boundary scattering is of determinative significance.

Some more attempts of theoretical description of rough-wall resonator spectra can be noted. However, they all run
inevitably into great difficulties associated with the lack of appropriate methods for considering the wave scattering
at random inhomogeneous boundaries of the system under investigation. For instance, in spectral analysis of open-type
resonators possessing randomly rough boundaries, namely, dielectric disk resonators (DDRs), the method of polarization
currents (the so-called volume current method, VCM) was widely utilized in due time \cite{bib:Kuznetsov83}. Primarily, the method was invented to make plausible estimations of radiation loss of whispering gallery (WG) oscillations of randomly bounded open electrodynamic systems \cite{bib:LillleChu96,bib:LillleLaineChu97}.
This method reduces essentially to simulation of oscillation scattering produced by real boundary inhomogeneities by adding the fictitious sources (currents) randomly distributed near the unperturbed (nonrough) boundary of the system. However, in our recent paper, Ref.~\cite{bib:GanErTar09}, we put forward the arguments in favor of the inconsistency of using the VCM to describe the radiation loss in DDRs with rough boundaries and obtained the solution of this intricate problem with a more efficient and reliable method, namely, the direct solution of wave equations for fields in such systems. This was made possible due to elaboration of the specific method for resonance mode separation, which is applicable to systems with arbitrary degrees of disorder. We succeeded in showing that the decay in quality factors of
resonators with randomly rough boundaries results, basically, from the roughness-induced intermode scattering. Due to this type of scattering, the energy of high-$Q$ WG modes goes over to other, less stable, modes and in such a way radiates out of the resonator volume. Interestingly, the level of radiation loss of a randomly rough DDR turns out to be mainly governed by the average slope of the asperities (or, in other words, by their sharpness) and is much less sensitive to the amplitude of boundary fluctuations.

It should be noted that taking into account the random roughness of wave system boundaries is a rather complex mathematical problem, which has by now a long-lasting history. Basically, different methods for describing the wave scattering by such boundaries were previously developed for open and infinite systems, such as sea surface \cite{bib:BassFuks79,bib:Ogilvy91,bib:Voron95}. Even so, the number of problems subject to being completely solved was principally restricted by the applicability of the small-perturbation method (small amplitude and small slope of boundary asperities) and/or Kirchhoff approximation (the inclusion of one-fold scattering in the geometrical-optic approach).As to the essentially confined systems such as resonators
and quantum dots, their spatial boundedness imposes, on its own account, significant limitations on the possibility to
apply statistical methods for describing the wave scattering in such systems. Therefore, having in mind the essential
finiteness of the object of our study in this paper, the methods used previously to cover wave scattering caused by boundary roughness need serious modifications.

To calculate the spectra of resonance systems with randomly inhomogeneous boundaries we suggest a method which
makes it possible, above all, to perform calculations without imposing the substantial limitations on the random asperity amplitude. Second, the method enables one to avoid any restrictions regarding the asperity sharpness. The particular advantage of the suggested method is that it is not rigidly bounded to symmetries of the system under consideration.

The method includes two basic stages. Initially, the problem regarding the oscillations in cylindrical resonator with
arbitrarily rough walls is reduced, by means of the boundary conformal smoothing, to the problem for oscillations in the resonator whose boundaries are ideally homogeneous.\footnote{The idea of such a smoothing goes back to Migdal's work on resonance levels of deformed nuclei, see monograph~\cite{bib:Migdal75}} In this way, one is able to introduce a new complete set of eigenmodes (we call them gradient renormalized, or GR, modes) which serve as a good basic approximation for developing the appropriate perturbation theory.

At the next stage, the GR modes are entirely separated, much in the same way as it is done for standard integrable
systems. Remarkably, however, these specific modes can be separated regardless of the level of the asperity sharpness, the latter being the measure of scattering degree. The deduced functional expressions for eigenfrequencies of the disordered resonator enable one to carry out a detailed analysis of its spectral properties to any desirable accuracy.

The conclusions we have made from our considerations are as follows. First of all, at the amplitude of boundary
inhomogeneities, which is small as compared to resonator dimensions, the level of the disorder is specified by the
universal gradient parameter, namely, the average slope of the asperities, rather than by the relationships between the mean asperity amplitude, the asperity correlation length, and the wavelength of the excited oscillations. Depending on whether the asperities are smooth (i.e., their mean slope ratio is small as compared to unity) or sharp, the resonator behaves either as a weakly or a strongly disordered wave system.

Next, as opposed to resonators with the disordered bulk infill, the distribution of nearest-neighbor spacings in the
spectra of surface-inhomogeneous resonators acquires the features typical for Wigner distribution (which is conventionally associated with chaotic properties) only when the unitarity of wave operator is violated, i.e., if dissipation and/or openness channels are available in the system. In the absence of dissipation the spectrum is completely regular,with zero-width resonance lines, but offers essentially different correlation properties and resonance level densities in the domains of smooth and sharp boundary asperities.

The result of particular interest we have achieved in the present work is the relationship established between the
roughness-induced decoherence of resonance modes and the resonator dissipative properties. The resonance line broadening related to (Raleigh-type) mode scattering due to randomly rough boundaries, being additive with the broadening caused
by the dissipation per se, is found to be also proportional to the dissipation rate. If dissipation (or resonator openness) is nonexistent, so will the roughness-induced broadening be unavailable. This fact fundamentally differentiates resonators with boundary inhomogeneities, which belong to the class of billiard-type systems, from resonators randomly inhomogeneous in bulk.

\section{The choice of the model and the problem formulation}

\begin{figure}[h]
  \captionstyle{centerlast}
  \centering
  \scalebox{.7}[.6]{\includegraphics{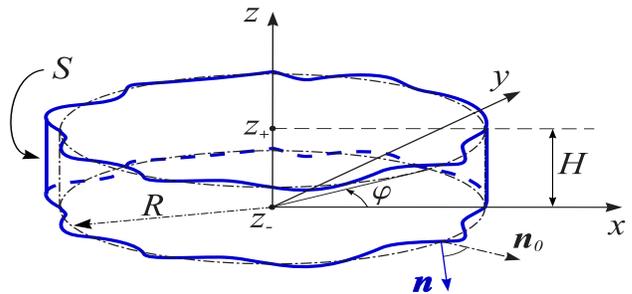}}
  \caption{Schematic view of cylindrical resonator with a randomly rough side boundary. Vectors $\mathbf{n}$ and $\mathbf{n_0}$ point toward outer normal to the rough and ideal boundary, respectively.    \hfill
 \label{fig1}}
\end{figure}

The object of our study is a resonator of cylindrical form with generatrix parallel to $z$-axis (see Fig.~\ref{fig1}), which is bounded by ideally-conducting plates passing perpendicular to $z$-axis through the points $z_{\pm}=\pm H/2$ and the ideally-conducting side boundary
\begin{equation}\label{SideBound}
  S=\left\{
  \begin{array}{lcl}
    r &=& R+\xi(\varphi)\\[4pt]
    \varphi &\in& [-\pi,\pi]
  \end{array}
  \right.\ .
\end{equation}
Here $R$ is the average radius of the resonator, i.\,e., the radius of an ideally circular cylinder whose volume coincides with that of the original rough cylinder having the same height.
The function $\xi(\varphi)$, which specifies inhomogeneity of the resonator side boundary, is assumed to be the random process with zero mean, $\big<\xi(\varphi)\big>=0$, and binary correlation function
\begin{equation}\label{BinCorr}
  \big<\xi(\varphi)\xi(\varphi')\big>=\sigma^2W(\varphi-\varphi')\ .
\end{equation}
Taking $\sigma$ in Eq.~\eqref{BinCorr} to be the mean-square height of the roughness, correlation function $W(\varphi)$ will be thought of as having unity maximum at $\varphi=0$ and decreasing to parametrically small values on the angle scale $|\Delta\varphi|=\varphi_c$. The correlation angle $\varphi_c$ is related to the roughness arc correlation length, $s_c$, through the obvious relationship $\varphi_c=s_c/R$. Both of the functions, $\xi(\varphi)$ and $W(\varphi)$, are taken to be periodic with the period~$2\pi$. In what follows, we regard the roughness as small scale in that the arc correlation dimension is small as compared to the characteristic size of the system,
\begin{equation}\label{Small-scale_roughness}
  s_c\ll R\ .
\end{equation}
Despite the resonator being a restricted system, this inequality suggests the conditions for spectrum self-averaging to be fulfilled with parametric accuracy. An additional argument in favor of the self-averaging is the well-known fact that classical dynamics in systems analogous to wave billiard under study, whose boundary pertains to the class of \emph{scattering} boundaries, is essentially ergodic \cite{bib:Birkhof31,bib:Hinchin41}.

Upon finding the Green's function $G(\mathbf{r},\mathbf{r}')$ of the Helmholtz equation inside the region depicted in Fig.~\ref{fig1}, by way of boundary conditions (BC) we use the equalities \cite{bib:Vainstein88}
\begin{subequations}\label{Bcond-rough-TE}
 \begin{equation}\label{sideBC}
   \frac{\partial\,G(\mathbf{r},\mathbf{r}')}{\partial\,r_n}\bigg|_{S}=0
 \end{equation}
and
 \begin{equation}\label{frontBC}
    G(\mathbf{r},\mathbf{r}')\bigg|_{z=\pm H/2}=0\ ,
 \end{equation}
\end{subequations}
which correspond to $H$-polarized electromagnetic oscillations (TE oscillations).\protect\footnote{The electric field of these oscillations is polarized parallel to the resonator butt-end plates, whereby these oscillations are more sensitive to fluctuations of the resonator side boundary than are the TM-type oscillations.} The derivative in Eq.~\eqref{sideBC} is taken over an outer normal line to actual, \emph{corrugated} side boundary (see~Fig.~\ref{fig1}).

In the general case the solution to wave equations with BC Eqs.~\eqref{Bcond-rough-TE} is a challenging task. In this paper, following the approach we have previously applied in Ref.~\cite{bib:GanErTar09}, we smooth out the rough boundary of the resonator to the ideal one, i.e., unperturbed by the asperities. In this way we manage to reduce the problem of wave scattering at the resonator actual boundary to the scattering by fluctuations of the effective medium filling
the bulk of a fictitious resonator, the cross section of which is the ideal circle.

We relate the coordinates in the initial cylinder system coupled to the resonator axis to new cylindric coordinates
(we mark them by the tilde sign) through the conformal transformation
\begin{eqnarray}
    \widetilde{r} &=& \frac{r}{1+\xi(\varphi)/R}\ ,\notag \\
\label{coord_transform}
    \widetilde{\varphi} &=& \varphi\ , \\\notag
    \widetilde{z} &=& z\ ,
\end{eqnarray}
whose Jacobian is equal to $1/\left[1+\xi(\varphi)/R\right]$. In the new coordinate system the Green function equation assumes the form
\begin{widetext}
\begin{equation}\label{GreenEqMain}
  \left(\frac{1}{r}\frac{\partial}{\partial r}r\frac{\partial}{\partial r}+
  \frac{1}{r^2}\frac{\partial^2}{\partial\varphi^2}+
  \frac{\partial^2}{\partial z^2}+K^2
  -\hat{V}^{(h)}-\hat{V}^{(s)}\right)
  G(r,\varphi,z;r'\varphi',z')=\frac{1}{r}\delta(r-r')\delta(\varphi-\varphi')\delta(z-z')
\end{equation}
\end{widetext}
(the tilde signs are omitted from here on), differing from the initial equation by the additional terms in the wave operator, namely, $\hat{V}^{(h)}$ and $\hat{V}^{(s)}$, which we call below the \emph{effective potentials},
\begin{subequations}\label{Potentials}
\begin{align}\label{heightPot}
  \hat{V}^{(h)} =& -\left(k^2+\frac{\partial^2}{\partial z^2}\right)
  \left[\beta^2(\varphi)-1\right]\ , \\
\label{slopePot}
  \hat{V}^{(s)} =& \left[\frac{\xi'(\varphi)}{R\beta(\varphi)}
  \frac{\partial}{\partial\varphi}+\frac{\partial}{\partial\varphi}
  \frac{\xi'(\varphi)}{R\beta(\varphi)}\right]\frac{1}{r}\frac{\partial}{\partial r}
 \notag\\
  &\phantom{\frac{\xi'(\varphi)}{R\beta(\varphi)}
  \frac{\partial}{\partial\varphi}+++}
  -\left[\frac{\xi'(\varphi)}{R\beta(\varphi)}\right]^2
  \frac{1}{r}\frac{\partial}{\partial r}r\frac{\partial}{\partial r}\ .
\end{align}
\end{subequations}
The analytical structure of these potentials is determined by the roughness function itself, $\xi(\varphi)$, and by its derivative $\xi'(\varphi)$. In Eq.~\eqref{GreenEqMain}, the notation $K^2=k^2-i/\tau_{\text{dis}}$ is introduced, where $k=\omega/c$ and $1/\tau_{\text{dis}}$ is the effective dissipation rate which we put in phenomenologically [time dependence of the fields is chosen to be of the form $\exp(-i\omega t)$]. Factor $\beta(\varphi)$ in Eqs.~\eqref{Potentials} is used to simplify the formula,
\begin{equation}\label{beta-func}
  \beta(\varphi)=1+\frac{\xi(\varphi)}{R}\ .
\end{equation}
Upper indices ``$h$'' and ``$s$'' at the potentials \eqref{Potentials} symbolize the fact that one of them, namely, the former, is governed by fluctuations of the asperity \emph{height} solely [i.\,e., directly by function $\xi(\varphi)$] whereas the potential $\hat{V}^{(s)}$ is mainly determined by the asperity \emph{slope} [by the derivative $\xi'(\varphi)$]. We consider $\xi(\varphi)$ to have a relatively small statistical value and put $\beta(\varphi)\thickapprox 1$ below.

In view of inequality \eqref{Small-scale_roughness}, one can neglect the correlation between functions  $\xi(\varphi)$ and $\xi'(\varphi)$ at angle distances larger than $\varphi_c$. Owing to this we can relate two mutually independent
physical mechanisms with potentials, Eqs.~\eqref{Potentials}, which are responsible for wave scattering by surface roughness, namely, the``height'' and the ``gradient'' mechanism, both predicted previously in Ref.~\cite{bib:MakTar-01}.

Note that contrary to the common belief (see, e.\,g., Ref.~\cite{bib:BassFuks79}), even though we hold the asperities to be small in height in the sense of inequality
\begin{equation}\label{SmallHeight}
  \sigma\ll R\ ,
\end{equation}
the effective potential \eqref{slopePot} may take on quite large values. The local value of this potential and, consequently, of the entire potential in Eq.~\eqref{GreenEqMain}, is determined by ratio $\xi'(\widetilde{\varphi})/R$, which under constraint \eqref{SmallHeight} assumes arbitrary large values. Clearly, this does not necessarily signify large scattering strength, as the latter is the integral rather than the local property of the potential. Below we formulate the particular parametric conditions enabling one to regard the effect of scattering due to effective potentials \eqref{Potentials} as either weak or strong.

\section{Gradient renormalization of wave operator and separation of oscillation modes}

In contrast to ``hight'' potential $\hat{V}^{(h)}$, ``slope'' potential $\hat{V}^{(s)}$ in Eq.~\eqref{GreenEqMain}, keeping in mind the prospective usage of perturbation theory, is defined in an awkward fashion as its average is not equal to zero. By separating this average as the operator summand,
\begin{equation}\label{Av_V(s)}
  \Av{\hat{V}^{(s)}}=
  -\Xi^2\frac{1}{r}\frac{\partial}{\partial r}r\frac{\partial}{\partial
  r}\ ,
\end{equation}
where we have introduced the gradient parameter
\begin{equation}\label{Xi-def}
  \Xi^2=\frac{1}{R^2}
  \Av{\left[\xi'(\varphi)\right]^2}
  \sim\frac{\sigma^2}{s_c^2}
\end{equation}
which specifies the asperity sharpness degree, we can recast equation \eqref{GreenEqMain} in the form
\begin{widetext}
\begin{equation}\label{GreenEq-alt}
  \bigg[(1+\Xi^2)\frac{1}{r}\frac{\partial}{\partial r}r
  \frac{\partial}{\partial r}+
  \frac{1}{r^2}\frac{\partial^2}{\partial\varphi^2}+
  \frac{\partial^2}{\partial z^2}+K^2
  -\hat{V}^{(h)}-\hat{V}^{(s1)}-\hat{V}^{(s2)}\bigg]
  G(r,\varphi,z;r'\varphi',z')=\frac{1}{r}\delta(r-r')\delta(\varphi-\varphi')\delta(z-z')
  \ .
\end{equation}
\end{widetext}
In Eq.~\eqref{GreenEq-alt}, two different slope potentials are introduced instead of the potential \eqref{slopePot}, whose configurational averages are by definition equal to zero, namely,
\begin{subequations}\label{SlopePots-defs}
\begin{align}
 \label{SlPot-1}
  \hat{V}^{(s1)} =& \frac{1}{R}\left[\xi'(\varphi)
  \frac{\partial}{\partial\varphi}+\frac{\partial}{\partial\varphi}
  \xi'(\varphi)\right]\frac{1}{r}\frac{\partial}{\partial r}\ ,\\
 \label{SlPot-2}
  \hat{V}^{(s2)} =& -\left[{\xi'}^2(\varphi)/R^2-\Xi^2\right]
  \frac{1}{r}\frac{\partial}{\partial r}r\frac{\partial}{\partial
  r}\ .
\end{align}
\end{subequations}
The function $\beta(\varphi)$, subject to proviso \eqref{SmallHeight}, is set equal to unity herein.

Let us proceed to Fourier representation using thereto a complete set of eigenfunctions of the differential operator in Eq.~\eqref{GreenEq-alt}, which are consistent with boundary conditions \eqref{Bcond-rough-TE} and $\varphi$-periodicity of the desired solution. In Eq.~\eqref{GreenEq-alt} with no random potentials, the variables are readily separated, thereby giving the following complete set of normalized eigenfunctions:
\begin{subequations}\label{Eigenfuncs}
  \begin{align}
  \label{Eigenfunc_123}
    &\ket{\mathbf{r};\bm{\mu}} =
    \ket{\varphi;n}\ket{r;l}^{(\widetilde{n})}\ket{z;q} \\ &\phantom{m}\big(\widetilde{n}=n/\sqrt{1+\Xi^2}\big)\ ;\notag \\[.5\baselineskip]
  \label{Eigenfunc_1}
    &\ket{\varphi;n} =(2\pi)^{-1/2}\mathrm{e}^{in\varphi}\\
    &\phantom{m}(n =0,\pm 1,\pm 2,\ \ldots) \ ,\notag \\
  \label{Eigenfunc_2}
    &\ket{r;l}^{(\widetilde{n})}=B_l^{(\widetilde{n})}\frac{\sqrt{2}}{R}
    J_{|\widetilde{n}|}(\gamma_l^{(|\widetilde{n}|)}r/R) \qquad \\
    &\phantom{m}(l=1,2,\ \ldots)\ ,\notag \\
  \label{Eigenfunc_3}
    &\ket{z;q}=\sqrt{\frac{2}{H}}\sin\left[\left(\frac{z}{H}+\frac{1}{2}\right)\pi
    q\right] \\
    &\phantom{m}(q=1,2,\ \ldots)\ .\notag
  \end{align}
\end{subequations}
Here, $\gamma_l^{(|\widetilde{n}|)}$ denote the positive zeros of function $J'_{|\widetilde{n}|}(t)$ enumerated by natural numbers ($l$) in ascending order. The normalization constant in ``radial'' eigenfunction \eqref{Eigenfunc_2} is
\begin{equation}\label{Bln-TE}
  B_l^{(\widetilde{n})}=\gamma_l^{(|\widetilde{n}|)}
  \left[\left(\gamma_l^{(|\widetilde{n}|)}\right)^2-\widetilde{n}^2\right]^{-1/2}
  \left[J_{|\widetilde{n}|}(\gamma_l^{(|\widetilde{n}|)})\right]^{-1}\ .
  \end{equation}

Below we refer to the Hermitian differential operator in Eq.~\eqref{GreenEq-alt}, which possesses the set of eigenfunctions \eqref{Eigenfuncs}, as the GR Laplace operator. In representation of this operator eigenmodes Eq.~\eqref{GreenEq-alt} reduces to a set of non-homogeneous coupled equations against the Green's function Fourier components,
\begin{equation}\label{Mode_Green}
  \left(K^2-\varkappa^2_{\bm{\mu}}-{\mathcal V}_{\bm{\mu}}\right)
  G_{\bm{\mu}\bm{\mu}'}-
  \sum_{\bm{\nu}\neq\bm{\mu}}{\mathcal
  U}_{\bm{\mu}\bm{\nu}}G_{\bm{\nu}\bm{\mu}'}=
  \delta_{\bm{\mu}\bm{\mu}'}\ .
\end{equation}
Here,
\begin{equation}\label{ModeGreen_matrix-def}
  G_{\bm{\mu}\bm{\mu}'}=\iint_{\Omega}d\mathbf{r}d\mathbf{r}'\bra{\mathbf{r};\bm{\mu}}
  G(\mathbf{r},\mathbf{r}')\ket{\mathbf{r}';\bm{\mu}'}
\end{equation}
is the Green's function in momentum representation, and $\Omega$ is the domain in coordinate space $\mathbb{R}^3$ occupied by the resonator with an unperturbed boundary. The quantity $\varkappa^2_{\bm{\mu}}$, which is the GR Laplasian eigenvalue, in what follows is called an \emph{unperturbed energy} of the $\bm{\mu}$th resonator mode,
\begin{equation}\label{mode_en}
  \varkappa^2_{\bm{\mu}}=
  (1+\Xi^2)\left(\frac{\gamma_l^{(|\widetilde{n}|)}}{R}\right)^2
  +\left(\frac{\pi q}{H}\right)^2\ .
\end{equation}
Accordingly, the eigenfunctions \eqref{Eigenfuncs} will be symbolically termed as unperturbed eigenfunctions, even though they contain gradient parameter $\Xi$.

Matrix $\|{\mathcal U}_{\bm{\mu}\bm{\nu}}\|$ of the Eq.~\eqref{Mode_Green} coefficients is composed of matrix elements of the total perturbation potential in Eq.~\eqref{GreenEq-alt}, which contains both the height potential and the gradient ones [$\hat{V}^{(pert)}(\mathbf{r})={{V}^{(h)}+\hat{V}^{(s1)}+\hat{V}^{(s2)}}$],
\begin{equation}\label{U_mode}
  {\mathcal U}_{\bm{\mu}\bm{\nu}}=\int_\Omega\textit{d}\,\mathbf{r}
  \bra{\mathbf{r};\bm{\mu}}\hat{V}^{(pert)}(\mathbf{r})\ket{\mathbf{r};\bm{\nu}}\
  .
\end{equation}
The diagonal element of this matrix, ${\mathcal U}_{\bm{\mu}\bm{\mu}}\equiv
{\mathcal V}_{\bm{\mu}}$, is deliberately separated from the sum in Eq.~\eqref{Mode_Green}. The scattering induced by this element is specified below as the intramode, or \emph{coherent}, scattering, and the potential ${\mathcal V}_{\bm{\mu}}$ will be named the intramode potential. On the contrary, all the constituents of matrix $\|{\mathcal U}_{\bm{\mu}\bm{\nu}}\|$ with distinct mode indices ($\bm{\mu}\neq\bm{\nu}$) are thought of as responsible for \emph{incoherent} scattering between different modes and are nominally referred to as intermode potentials.
In Refs.~\cite{bib:Tar00,bib:GanErTar09,bib:Tar05} it was shown that such a segregation of the whole set of potentials into ``coherent'' and ``incoherent'' subsets enables one to correctly deduce the whole set of uncoupled equations for diagonal elements of the entire Green's matrix $\|G_{\bm{\mu}\bm{\nu}}\|$, through which all other elements are then expressed by linear operation. Basically, this fact makes it easy to analyze the quantum (wave) systems subjected to
quite arbitrary external potentials, including nonlocal ones.

Equation \eqref{Mode_Green} allows the peculiar separation of variables provided that at least one of two fundamental symmetries, either $\mathcal{P}$ or $\mathcal{T}$, is broken. The variables are separated through the operator procedure detailed in Ref.~\cite{bib:GanErTar07}. The key point of the procedure is to introduce a certain trial mode propagator $G_{\bm{\nu}}^{(V)}$ for each of the harmonics, say, with mode index $\bm{\nu}$. This propagator represents the ``seed'' mode Green's function, i.\,e. the solution to Eq.~\eqref{Mode_Green} obtained without regard to intermode potentials, and, hence, intermode scattering,
\begin{equation}\label{G_trial}
  G_{\bm{\nu}}^{(V)}=\left[K^2-\varkappa^2_{\bm{\nu}}-{\mathcal
  V}_{\bm{\nu}}\right]^{-1}\ .
\end{equation}
Starting from this trial function, all the intermode propagators $G_{\bm{\nu}\bm{\mu}}$ (with $\forall\bm{\nu}\neq\bm{\mu}$) are expressed in terms of just one intramode Green's function $G_{\bm{\mu}\bm{\mu}}$ by means of functional equality:
\begin{equation}\label{Gnu->Gmumu}
  G_{\bm{\nu}\bm{\mu}}=\bm{P}_{\bm{\nu}}
  (\openone-\hat{\mathsf R})^{-1}\hat{\mathsf R}
  \bm{P}_{\bm{\mu}}G_{\bm{\mu}\bm{\mu}}\ .
\end{equation}
Here, operator $\hat{\mathsf R}$ has the meaning of a mode-mixing operator, whose domain of definition is mode subspace ${\mathsf{\overline M}_{\bm{\mu}}}$ containing the whole set of mode indices save index $\bm{\mu}$; $\bm{P}_{\bm{\mu}}$ is the projection operator that assigns the given value $\bm{\mu}$ to the nearest mode index of any operator standing next to it, no matter if it is to the left or to the right. The operator $\hat{\mathsf R}$ is expressed in the product form,
\begin{equation}\label{R-oper}
  \hat{\mathsf R}=\hat{\mathcal{G}}^{(V)}\hat{\mathcal U}\ ,
\end{equation}
where operators $\hat{\mathcal{G}}^{(V)}$ and $\hat{\mathcal U}$ are specified on ${\mathsf{\overline M}_{\bm{\mu}}}$ by their matrix elements
\begin{subequations}\label{G(V)U}
\begin{align}\label{GV-matr}
  & \bra{\bm{\nu}}\hat{\mathcal G}^{(V)}\ket{\bm{\nu}'}=
  G^{(V)}_{\bm{\nu}}\delta_{\bm{\nu}\bm{\nu}'}\ ,\\
  \label{U-matr}
  & \bra{\bm{\nu}}\hat{\mathcal U}\ket{\bm{\nu'}} =
  {\mathcal U}_{\bm{\nu}\bm{\nu'}}\ .
\end{align}
\end{subequations}

Putting then mode index $\bm{\mu}'=\bm{\mu}$ in Eq.~\eqref{Mode_Green} and substituting all the intermode propagators in the form \eqref{Gnu->Gmumu}, we arrive at the infinite set of \emph{uncoupled} equations for all intramode propagators $G_{\bm{\mu}\bm{\mu}}$, whose solution is
\begin{equation}\label{Gmumu-last}
  G_{\bm{\mu}\bm{\mu}}=\left[K^2-\varkappa^2_{\bm{\mu}}-{\mathcal
  V}_{\bm{\mu}}-{\mathcal T}_{\bm{\mu}}\right]^{-1}
  \qquad \text{(for $\forall\bm{\mu}$'s)}\ .
\end{equation}
Here,
\begin{equation}\label{T-oper}
  {\mathcal T}_{\bm{\mu}}=\bm{P}_{\bm{\mu}}\hat{\mathcal U}
  (\openone-\hat{\mathsf R})^{-1}\hat{\mathsf R}\bm{P}_{\bm{\mu}}
\end{equation}
is the portion of the $\bm{\mu}$th mode eigenenergy which is related to the intermode scattering only. The whole set of intramode propagators, Eq.~\eqref{Gmumu-last}, in conjunction with relationship Eq.~\eqref{Gnu->Gmumu}, solves entirely the perturbed resonator Green's function. In such a way we have reduced the problem of determining the randomly rough resonator spectrum to finding the poles of solely diagonal elements of matrix $\|G_{\bm{\mu}\bm{\nu}}\|$.

From Eq.~\eqref{Gmumu-last} it can be easily seen that in the general case all eigenfrequencies of the resonator we study appear to be complex valued. The complexity is apparently introduced by two sources. One of them is related to dissipation directly, being accounted for through the dissipation rate entering complex "energy" of the resonator, $K^2=k^2-i/\tau_{\text{dis}}$. The other originates from the $T$ potential \eqref{T-oper}, which is nothing but the diagonal (in mode representation) element of the regularized T matrix well-known in quantum scattering theory \cite{bib:Newton68,bib:Taylor72}. This potential functional structure suggests that it can be interpreted as taking into account the effective \emph{interaction} between the particular mode $\bm{\mu}$ and the rest of trial modes with $\forall\bm{\nu}\neq\bm{\mu}$. 

Eventually, with the account of intermode interaction, the mode $\bm{\mu}$ becomes either coherent or incoherent, depending on whether the dissipation is present in the resonator ($1/\tau_{\text{dis}}\neq 0$) and/or the potential \eqref{T-oper} contains an imaginary part. The latter kind of imaginarity comes about in the self-energy of the particular wave or quantum system (in our case, the mode state $\bm{\mu}$) interacting with the hardly controllable ``environment'' if one tries to describe this system separately. Normally the imaginary addendum to the isolated system self energy is interpreted as environment-induced indefiniteness of its energy levels, or, in other words, as the result of its decoherence \cite{bib:Zurek03} (see also Ref.~\cite{bib:PazZurek99}). Hereafter we also refer to the imaginary part (if any) of the $T$ potential \eqref{T-oper} as the decoherence rate of the $\bm{\mu}$th eigenlevel.

\section{Statistical analysis of the randomly rough resonator spectrum}

Since the steady-state propagation of waves in the resonator under study is expected to be ergodic (see Ref.~\cite{bib:Zaslavsky81} and references therein), one can obtain its spectrum by performing \emph{configurational} averaging of the Green's function~\eqref{Gmumu-last}. We denote this type of averaging by angular brackets, $\av{\cdot}$.

The net random potential in Eq.~\eqref{GreenEqMain} and, consequently, its matrix elements specifying the intra- and intermode potentials in Eq.~\eqref{Gmumu-last}, are strongly dependent on the statistical properties of the roughness function $\xi(\varphi)$. The intramode scattering produced by the potential ${\mathcal V}_{\bm{\mu}}$ may be regarded as either weak or strong, subject to small or not, in comparison with the unperturbed mode energy $\varkappa^2_{\bm{\mu}}$, appears to be the mean-square norm of this potential estimated on the unperturbed mode eigenfunction.

As far as intermode scattering is concerned, the associated additions to the unperturbed spectrum, Eq.~\eqref{mode_en}, result mainly from the potential ${\mathcal T}_{\bm{\mu}}$, whose average is not equal to zero. This potential has a rather involved functional structure; therefore it seems to be reasonable to characterize its effect by estimating the norm of operator $\hat{\mathsf R}$ which controls the trial mode intermixing. Note that these modes are not the
eigenstates of a parent nonrough resonator. Their energies, as seen from  Eq.~\eqref{mode_en}, contain the sharpness parameter $\Xi^2$, and, therefore, the field of each of these modes may be represented
as a bundle of parent, nonrough resonator modes. Keeping this fact in mind, we call the above-introduced trial modes, which are specified by eigenenergies \eqref{mode_en}, the unperturbed \emph{composite} modes.

\subsection{The comparative estimation of scattering mechanisms for composite modes}

\subsubsection{Intramode scattering}%
\label{Intramode_scat-estim}

Under restriction \eqref{SmallHeight}, the height potential, Eq.~\eqref{heightPot}, can be represented with good accuracy by approximate expression
\begin{equation}\label{HeigtPot-appr}
  \hat{V}^{(h)}\approx -2\left(k^2+\frac{\partial^2}{\partial z^2}\right)
  \frac{\xi(\varphi)}{R}\ .
\end{equation}
Since the random process $\xi(\varphi)$ is supposed to be ergodic, the diagonal matrix element of potential \eqref{HeigtPot-appr}, taken between functions \eqref{Eigenfunc_1} and its complex conjugate, is equal to zero. Diagonal matrix elements of either of gradient potentials \eqref{SlopePots-defs} also vanish: (i) for the potential $\hat{V}^{(s1)}$ this is because of its periodicity in angle $\varphi$, and (ii) as regards the potential $\hat{V}^{(s2)}$ this is due to the assumed ergodicity of the roughness function.\footnote{This assumption is justified to the measure of inequality \eqref{Small-scale_roughness}.} Hence, under our assumptions regarding the statistical properties of $\xi(\varphi)$ there is no coherent scattering of composite modes which would be caused by random roughness of the resonator side boundary.

\subsubsection{Intermode scattering}

The effect of intermode scattering on the resonator spectrum, that is, on the pole structure of function \eqref{Gmumu-last}, is governed by the potential \eqref{T-oper}, which is defined as the diagonal matrix
element of the regularized $T$ matrix. The functional structure of this matrix and, consequently, the magnitude and the analytical structure of the potential ${\mathcal T}_{\bm{\mu}}$ are determined by the value of the mode-mixing operator $\hat{\mathsf R}$. We regard the intermode scattering as either weak or strong depending on whether the average norm of this operator is small or large as compared with unity. In view of the operator $\hat{\mathsf R}$ multiplicative structure, we will estimate this norm using the formula
\begin{equation}\label{R-norm_estim}
  \Av{\|\hat{\mathsf R}\|^2}=
  \max_{\bm{\mu}}\AV{\Big|\sum_{\bm{\nu}}G_{\bm{\nu}}^{(V)}
  {\mathcal U}_{\bm{\nu}\bm{\mu}}\Big|^2}\ .
\end{equation}
Here, trial Green function $G_{\bm{\nu}}^{(V)}$ is independent of the roughness function $\xi(\varphi)$, with regard to remarks made in the previous subsection. The averaging of the potential ${\mathcal U}_{\bm{\nu}\bm{\mu}}$ modulus square, which is in fact simple but requires a gret deal of tedious calculations, is detailed in the Appendix. Eventually, we obtain the following estimations:
\begin{subequations}\label{R_oper-norms}
\begin{align}
  & \Av{\|\hat{\mathsf R}^{(h)}\|^2}\sim
  \left(\sigma/R\right)^2\ ,
\label{R(h)_norm}\\[6pt]
  & \Av{\|\hat{\mathsf R}^{(s1)}\|^2}\sim
  \left(\sigma/s_c\right)^2\sim\Xi^2\ ,
\label{R(s1)_norm}\\[6pt]
  & \Av{\|\hat{\mathsf R}^{(s2)}\|^2}\sim
  \left(\sigma/R\right)^4\ .
\label{R(s2)_norm}
\end{align}
\end{subequations}
From these estimates one can infer that, in view of inequality \eqref{SmallHeight}, the potentials ${\mathcal U}_{\bm{\mu}\bm{\nu}}^{(h)}$ and ${\mathcal U}_{\bm{\mu}\bm{\nu}}^{(s2)}$ result in negligibly weak intermode scattering. For the scattering induced by the potential ${\mathcal U}_{\bm{\mu}\bm{\nu}}^{(s1)}$, it can be easily seen that under conditions \eqref{Small-scale_roughness} and \eqref{SmallHeight}, first, the norm of the corresponding term in the intermixing operator, ${\mathsf R}^{(s1)}$, is the uppermost among the norms presented in Eqs.~\eqref{R_oper-norms}, and second, the operator ${\mathsf R}^{(s1)}$ norm can take on both small and large values, as compared to unity. This enables us to analytically treat any of the limiting cases, both of weak and of strong intermode scattering. \\

\subsection{Evaluation of spectrum corrections}
\label{Weak_Scattering}

The composite mode spectrum given by Eq.~\eqref{mode_en}, even without additional corrections induced by potentials entering Eq.~\eqref{Mode_Green}, can differ essentially from the initial, unperturbed resonator spectrum. The difference is either small or large depending on whether the mean asperity slope (in fact, the parameter $\Xi^2$) is small or large as compared to unity. As may be seen from Eq.~\eqref{R(s1)_norm}, it is exactly the same parameter that governs the level of composite mode mixing. Below, while analyzing the spectral corrections, we keep in mind only the ones that result from the  scattering induced by the potential $\hat{V}^{(s1)}$.

\subsubsection{Weak scattering limit}

As it follows from estimates \eqref{R_oper-norms}, the intermode scattering, or, in other words, the entanglement of composite resonant modes, is weak provided that two inequalities hold simultaneously, namely,
\begin{equation}\label{Weak_intermode_scat}
\begin{aligned}
   & \sigma/R  \ll 1\ ,\\[3pt]
   & \sigma/s_c  \ll 1\ ,
\end{aligned}
\end{equation}
meaning the smallness (on average) of both the amplitude and the slope of random asperities. In this limiting case, one can expand the inverse operator in Eq.~\eqref{T-oper} in power series in the operator $\hat{\mathsf R}$, thereby obtaining the approximate expression
\begin{equation}\label{T_mu-weak_scat}
  {\mathcal T}_{\bm{\mu}}\approx \bm{P}_{\bm{\mu}}\hat{\mathcal
  U}\hat{G}^{(V)} \hat{\mathcal U}\bm{P}_{\bm{\mu}}=
  \sum_{\bm{\nu}\neq\bm{\mu}}{\mathcal U}_{\bm{\mu}\bm{\nu}}
  G^{(V)}_{\bm{\nu}}{\mathcal U}_{\bm{\nu}\bm{\mu}}
\end{equation}
for the diagonal element of the \emph{T}-matrix. The correlator of intermode potentials in Eq.~\eqref{T_mu-weak_scat}, subject to definition \eqref{BinCorr} and estimates \eqref{R_oper-norms}, is calculated as
\begin{align}
  \Av{{\mathcal U}_{\bm{\mu}\bm{\nu}}{\mathcal
  U}_{\bm{\nu}\bm{\mu}}}\approx &
  -\delta_{q_{\bm{\nu}}q_{\bm{\mu}}}
  \left(\frac{\sigma}{R}\right)^2
   \notag\\
 & \times\widetilde{W}(n_{\bm{\mu}}-n_{\bm{\nu}})
  \left(n^2_{\bm{\mu}}-n^2_{\bm{\nu}}\right)^2
  \mathcal{D}_{\bm{\mu}\bm{\nu}}
  \mathcal{D}_{\bm{\nu}\bm{\mu}}\ ,
\label{<U_munuUnumu>-weak}
\end{align}
where $\mathcal{D}_{\bm{\mu}\bm{\nu}}$ is the matrix element of the operator $\left(1/r\right)\left(\partial/\partial r\right)$, which is taken between eigenfunctions~\eqref{Eigenfunc_2},
\begin{equation}\label{D_munu}
  \mathcal{D}_{\bm{\mu}\bm{\nu}}=\int\limits_0^Rdr
  \bra{r;l_{\bm{\mu}}}^{(\widetilde{n}_{\bm{\mu}})}\frac{\partial}{\partial r}
  \ket{r;l_{\bm{\nu}}}^{(\widetilde{n}_{\bm{\nu}})}\ .
\end{equation}
By representing the average mass operator for mode $\bm{\mu}$ as a sum of real and imaginary parts, $\Av{{\mathcal T}_{\bm{\mu}}}={\Delta\varkappa_{\bm{\mu}}^2+i\big/\tau_{\bm{\mu}}^{(\text{sc})}}$, we arrive at the following expressions for resonance line shifting and broadening, which are associated with composite mode scattering,
\begin{subequations}\label{DeltaKappa+1/tau}
\begin{align}
\label{DeltaKappa}
   \Delta\varkappa_{\bm{\mu}}^2 & =
   \varkappa^4_{\bm{\mu}}\sum_{\bm{\nu}\neq\bm{\mu}}A_{\bm{\mu}\bm{\nu}}
   \frac{\varkappa_{\bm{\mu}}^2-\varkappa_{\bm{\nu}}^2}%
   {\big(\varkappa_{\bm{\mu}}^2-\varkappa_{\bm{\nu}}^2\big)^2
   +\big(1\big/\tau_{\text{dis}}\big)^2}\ ,\\[6pt]
   \frac{1}{\tau_{\bm{\mu}}^{(\text{sc})}} & =
   \varkappa^4_{\bm{\mu}}\sum_{\bm{\nu}\neq\bm{\mu}}A_{\bm{\mu}\bm{\nu}}
   \frac{1\big/\tau_{\text{dis}}}%
   {\big(\varkappa_{\bm{\mu}}^2-\varkappa_{\bm{\nu}}^2\big)^2
   +\big(1\big/\tau_{\text{dis}}\big)^2}\ .
\label{1/tau}
\end{align}
\end{subequations}
The weight factor $A_{\bm{\mu}\bm{\nu}}$ in Eqs.~\eqref{DeltaKappa+1/tau} is given by
\begin{widetext}
\begin{align}
  A_{\bm{\mu}\bm{\nu}}=-\delta_{q_{\bm{\nu}}q_{\bm{\mu}}}
  4 & \left(\frac{\sigma}{R}\right)^2\widetilde{W}(n_{\bm{\mu}}-n_{\bm{\nu}})
  \frac{\left(n^2_{\bm{\mu}}-n^2_{\bm{\nu}}\right)^2}{(\varkappa_{\bm{\mu}}R)^4}
  \left[B_{l_{\bm{\mu}}}^{(\widetilde{n}_{\bm{\mu}})}\right]^2
  \left[B_{l_{\bm{\nu}}}^{(\widetilde{n}_{\bm{\nu}})}\right]^2\notag
  \\
  & \times \int\limits_0^1dt
  J_{|\widetilde{n}_{\bm{\mu}}|}
  \left(\gamma_{l_{\bm{\mu}}}^{(|\widetilde{n}_{\bm{\mu}}|)}t\right)
  \frac{\partial}{\partial t}
  J_{|\widetilde{n}_{\bm{\nu}}|}
  \left(\gamma_{l_{\bm{\nu}}}^{(|\widetilde{n}_{\bm{\nu}}|)}t\right)
  \int\limits_0^1dt'
  J_{|\widetilde{n}_{\bm{\nu}}|}
  \left(\gamma_{l_{\bm{\nu}}}^{(|\widetilde{n}_{\bm{\nu}}|)}t'\right)
  \frac{\partial}{\partial t'}
  J_{|\widetilde{n}_{\bm{\mu}}|}
  \left(\gamma_{l_{\bm{\mu}}}^{(|\widetilde{n}_{\bm{\mu}}|)}t'\right)\
  .
\label{A_(munu)}
\end{align}
\end{widetext}

Interestingly, this factor is not positive definite in the general case. If we consider, instead of TE polarized oscillations, the oscillations of TM type, for which boundary conditions are applicable
\begin{equation}\label{sideBC-TM}
  G(\mathbf{r},\mathbf{r}')\big|_{S}=0
  \qquad\text{and}\qquad
  \frac{\partial\,G(\mathbf{r},\mathbf{r}')}{\partial z}\bigg|_{z=\pm H/2}=0
\end{equation}
rather than BC~Eq.~\eqref{Bcond-rough-TE} (see Ref.~\cite{bib:Vainstein88}), then the weight factors appropriate to the transitions between any two different modes would be necessarily positive. This stems from the fact that for TM oscillations the quantities $\gamma_{l}^{(|\widetilde{n}|)}$ in Eq.~\eqref{A_(munu)} would be zeros of the Bessel function itself rather than zeros of its derivative. Based on this fact, antisymmetry of the tensor $\mathcal{D}_{\bm{\mu}\bm{\nu}}$ would follow and, correspondingly, the antisymmetry of each of the integrals in Eq.~\eqref{A_(munu)}. As for TE oscillations being considered here, the sign of the coefficient $A_{\bm{\mu}\bm{\nu}}$ for different pairs of mode indices is not \emph{a~priori} definite. This implies that the intermode scattering contribution to \emph{Q} factors and shifting of different lines can be of different sign. To verify this anticipation by analytical means seems to be an unrealistic task. Therefore we subsequently analyze the rough-resonator spectrum numerically, using for this purpose  Eqs.~\eqref{DeltaKappa+1/tau} and \eqref{G_mumu-strong_scat_limit}.

Yet, even at this stage we are in a position to make some qualitative statements regarding the influence of surface
inhomogeneities on the resonator \emph{Q} factor as well as on the position of resonance lines. Assuming the azimuth mode indices to be large as compared to unity, one can replace exact zeros $\gamma_{l}^{(|\widetilde{n}|)}$ with their asymptotic expressions (see Ref.~\cite{bib:AbrSteg64}). For scattering-induced decay rate \eqref{1/tau}, the approximative formula readily follows
\begin{widetext}
\begin{equation}\label{SlopetWidth-weak_2}
  \frac{1}{\tau_{\bm{\mu}}^{(\text{sc})}}\sim
  \frac{1}{\tau_{\text{dis}}}
  \left(\frac{\sigma}{R}\right)^2
  \sum_{n_{\bm{\nu}}}(n^2_{\bm{\mu}}-n^2_{\bm{\nu}})^2
  \widetilde{W}(n_{\bm{\mu}}-n_{\bm{\nu}})
  \sum_{l_{\bm{\nu}}}
  \frac{1}{\left(l_{\bm{\mu}}-l_{\bm{\nu}}
  +\frac{|\widetilde{n}_{\bm{\mu}}|-|\widetilde{n}_{\bm{\nu}}|}{2}\right)^2+\big(R^2/\tau_{\text{dis}}\big)^2}\ .
\end{equation}
\end{widetext}
Here, the sum over radial mode indices, subject to the restriction imposed on the summation region in Eqs.~\eqref{DeltaKappa+1/tau}, is evaluated to be of order $1+\big(R^2/\tau_{\text{dis}}\big)$. The convergence of
the sum over azimuth indices is provided by the correlation function $\widetilde{W}(n_{\bm{\mu}}-n_{\bm{\nu}})$, from which we obtain %
\begin{equation}\label{SlopetWidth-weak_3}
  \frac{1}{\tau_{\bm{\mu}}^{(\text{sc})}}\sim
  \frac{1}{\tau_{\text{dis}}}\Xi^2\big(n_{\bm{\mu}}+\varphi_c^{-1}\big)^2\ .
\end{equation}

Because each of the resonance line \emph{Q} factors is determined by the sum of the dissipation rate and the rate of roughness-induced attenuation,
\begin{equation}\label{Q-factor}
  Q_{\bm{\mu}}^{-1}=
  \frac{\varkappa_{\bm{\mu}}}{\sqrt{1/\tau_{\text{dis}}+1/\tau_{\bm{\mu}}^{(\text{sc})}}}\ ,
\end{equation}
the emergence of random inhomogeneities on the resonator side boundary may result either in the preservation of the
specific line width, if the condition ${\Xi^2\big(n_{\bm{\mu}}+\varphi_c^{-1}\big)^2\lesssim 1}$ holds
true, or in its significant broadening, with this inequality changing to the opposite one. It is, however, intriguing that \emph{Q} factors of all resonance lines, subject to inequality \eqref{SmallHeight}, remain
infinite in the absence of dissipation, irrespective of whether the resonator side boundary is rough or not. At the same time, the positioning of these lines can differ considerably from that pertinent to the initial, nonrough resonator. Even if the asperities may be very smooth ($\Xi\ll 1$), the shift of the specific line, being determined by the difference between energy \eqref{mode_en} and the analogous energy calculated for $\Xi$ exactly equal to zero, can substantially exceed the mean level spacing. In principle, this single fact suffices to change significantly
the distribution of resonances along the frequency axis, as compared to the inhomogeneity-free resonator. But apart from that, the additional shift should also be taken into account, which is governed by Eq.~\eqref{DeltaKappa}, whose relative value, as numerical estimations do reveal, may significantly exceed the parameter $\Xi^2$ in view of a large number of terms in the sum.

Of dominant importance, however, is the fact that the spectrum of weakly rough [in the sense of inequalities \eqref{Weak_intermode_scat}] resonator remains strictly discrete (i.\,e., having exactly zero line width) in the entire absence of dissipation [see Eq.~\eqref{1/tau}] despite the quite significant intermode scattering. This entirely correlates with a common knowledge regarding the spectra of conservative ballistic systems \cite{bib:Birkhof31,bib:Hinchin41}. So far as the wave equations for the resonator in question prove to be separable despite its boundaries' complex structure, one may (and should) regard such a resonator as the exactly integrable system. In Sec.~\ref{SpestrumStatistics},  we provide the results of numerical calculations which enable one
to make some definite conclusions of a general type about the chaotic properties of spectra of wave systems whose classical analogs are nonintegrable.

\subsubsection{Strong scattering limit}
\label{Strong_Scattering}

By strong intermode scattering we imply the limiting case pertinent to the large (as compared to unity) norm of the
mode-mixing operator $\hat{\mathsf R}$ in Eq.~\eqref{T-oper}. Subject to strong mode mixing, each of the modes in this case becomes quasiuniformly distributed over a large region in mode space.\footnote{In classical chaos theory the characteristic size of this region is specified as the \emph{dynamic localization length} \cite{bib:ChirShep86,bib:ChirIzrShep88}.} For the resonator we study here this corresponds (see Eq.~\eqref{R(s1)_norm}) to inequality
\begin{equation}\label{Strong_scat-condition}
  \Xi^2\sim\left(\sigma/s_c\right)^2\gg 1\ .
\end{equation}
The asperities that obey this condition are referred to as ``sharp'' boundary inhomogeneities.


In the limiting case \eqref{Strong_scat-condition}, the operator potential \eqref{T-oper} may be expanded in a power series of the inverse operator~$\hat{\mathsf R}^{-1}$. The expression located between projection operators in Eq.~\eqref{T-oper} is then transformed to
\begin{align}\label{T_matr-large_R}
  \hat{\mathcal U}
  \big(\openone-\hat{\mathsf{R}}\big)^{-1}\hat{\mathsf{R}}
  \approx -\hat{\mathcal U}-\hat{\mathcal{G}}^{(V)-1}-
  \hat{\mathcal{G}}^{(V)-1}\hat{\mathcal
  U}^{-1}\hat{\mathcal{G}}^{(V)-1}\ .
\end{align}
Here, the symbol $\hat{\mathcal{G}}^{(V)}$ denotes the diagonal (in mode representation) operator on ${\mathsf{\overline M}_{\bm{\mu}}}$ whose matrix elements are trial Green's functions, Eq.~\eqref{G_trial}, with all indices $\bm{\nu}\neq\bm{\mu}$. Substituting expansion \eqref{T_matr-large_R} into potential \eqref{T-oper} we obtain the following asymptotic expression for the diagonal mode propagator,
\begin{align}\label{G_mumu-strong_scat_limit}
  G_{\bm{\mu}\bm{\mu}}\approx & \frac{G^{(V)}_{\bm{\mu}}}{2}
  \left[1+\frac{1}{2}\big(
  \hat{\mathsf{R}}^{-1}\big)_{\bm{\mu}\bm{\mu}}\right]^{-1} \notag\\
 & =
  \frac{G^{(V)}_{\bm{\mu}}}{2}
  \left[1+\frac{1}{2}{G^{(V)}_{\bm{\mu}}}^{-1}\big(
  \hat{\mathcal{U}}^{-1}\big)_{\bm{\mu}\bm{\mu}}\right]^{-1}\ .
\end{align}
The second term in square brackets in Eq.~\eqref{G_mumu-strong_scat_limit} is a small (in the parameter ${\|\hat{\mathsf R}\|^{-1}\ll 1}$) correction to the first one. Moreover, this corrective summand turns out to be less the closer the external frequency $\omega$ is to the resonance frequency of the ``unperturbed'' $\bm{\mu}$th resonator harmonics [the
latter corresponding to vanishing the real part of ${G^{(V)}_{\bm{\mu}}}^{-1}$]. This implies that strong intermode scattering caused by the resonator boundary roughness does not have a great impact upon the spectral energies' analytical structure against the weak scattering situation. In both limiting cases the resonance frequencies are specified to a good parametric accuracy by the poles of the trial Green's function $G^{(V)}_{\bm{\mu}}$, i.\,e., they can be found from the equality $k\approx\varkappa_{\bm{\mu}}$.

At first glance, the width of the resonance lines behaves in a somewhat abnormal fashion. In the absence of random inhomogeneities on the resonator side boundary it is determined by dissipation loss only, being equal, by the order of magnitude, to the rate $1/\tau_{\text{dis}}$ (in units of ``energy'' $k^2$). On sharpening the asperities within the range $\Xi\lesssim 1$ the resonance line \emph{Q} factors decrease as their widths are defined by the sum of the dissipation rate and the rate of scattering-induced mode decoherence.


On passing to the region $\Xi>1$, as the notions of ray dynamics suggest, an increase of chaotic nature should be
observed, which inevitably changes over to the ergodic stage. In the mode language this would imply the rearrangement of the excited mode energy between a large number of resonator modes and, as a consequence, the quasihomogeneous infilling
of the resonator volume by the electromagnetic field, irrespective of the field structure of the mode supposedly excited by the source.

As a matter of fact, the result of the wave equation solution is different. Specifically, the resonator spectrum is
back to its original, formally undisturbed form\eqref{mode_en} when the asperities are gradually sharpened. As to the disorder-induced composite mode decay rate [the analog of the weak-scattering formula Eq.~\eqref{1/tau}], with infinite sharpening of the asperities it progressively vanishes. In the limit of $\Xi\to\infty$ the width of resonance lines returns to the initial state, being governed by dissipation loss only.

Such peculiar behavior of resonance line width cannot be interpreted in terms of classical, i.\,e. trajectory-based,
chaos theory as the classic analog of the system we study here is definitely the {\emph{K} system} \cite{bib:Bunimovich74}. A decrease in the scattering-related portion of the resonance line width, as the random component of the resonator boundary increases, may solely be explained by the interference of scattered modes, whose contribution is lowered (according to the random-phase summation principle) with an increase in the number of interfering modes.

\section{Numerical analysis of the rough-resonator spectrum}
\label{SpestrumStatistics}

Now we present the results of numeric analysis of spectral properties of the rough resonator in the study, using the
analytical formulas obtained in the previous section. The subject of the analysis is the statistical distribution of nearest level spacings (NLS) conventionally defined through unfolded spacings $s_i$,
\begin{equation}\label{Spacing-def}
  s_i\approx\left(E_{i+1}-E_i\right)g(E_i)\ ,
\end{equation}
where $g(E_i)$ is the mean density of states evaluated at energy $E_i$.

According to the long-standing conjecture of Berry and Tabor \cite{bib:BerryTabor77}, the spectrum of an integrable system (normally referred to as a regular spectrum) is described by the NLS distribution following Poisson's law, $p(s)=\exp(-s)$. This distribution is normally associated with the absence of level correlation and, as a consequence, with their clusterization near $s=0$. If there exists some chaotic component in the
spectrum, the levels become correlated to a certain extent and subjected to a kind of repulsion, which is manifested
through the gradual transformation of $p(s)$ from Poissonian- to Wigner-type distribution, $p(s)=(\pi/2)s\exp(-\pi s^2/4)$. The latter distribution is typical for systems with extremely chaotic classical dynamics~\cite{bib:Mehta67}.

One can infer from our results, Eqs.~\eqref{DeltaKappa+1/tau} and \eqref{G_mumu-strong_scat_limit}, which
apply to weak and strong intermode scattering, respectively, that in the idealized dissipation-free resonance system the line width should be zero, whatever the strength of the seed-mode scattering induced by surface inhomogeneities. Yet, the mere presence of this type of scattering, even though it is weak, results in substantial redistribution of resonance frequencies in comparison with the case where boundary asperities are entirely absent. The shift of each resonance line from its starting position (at $\Xi=0$) is determined by the sum of seed energy \eqref{mode_en} and the correction to this energy resulting from composite mode scattering.
In the smooth roughness domain corresponding to $\Xi^2\ll 1$ the correction to energy, Eq.~\eqref{mode_en}, has the form of Eq.~\eqref{DeltaKappa}. In the opposite limiting case, where asperities are strongly sharp ($\Xi^2\gg 1$), we regard
this correction as negligibly small, using expression \eqref{G_mumu-strong_scat_limit} with no corrective term for the $\bm{\mu}$th mode trial Green's function, $G_{\bm{\mu}}^{(V)}$.

\begin{figure*}[h!!]
  \captionstyle{centerlast}
  \centering
  \scalebox{.35}[.35]{\includegraphics{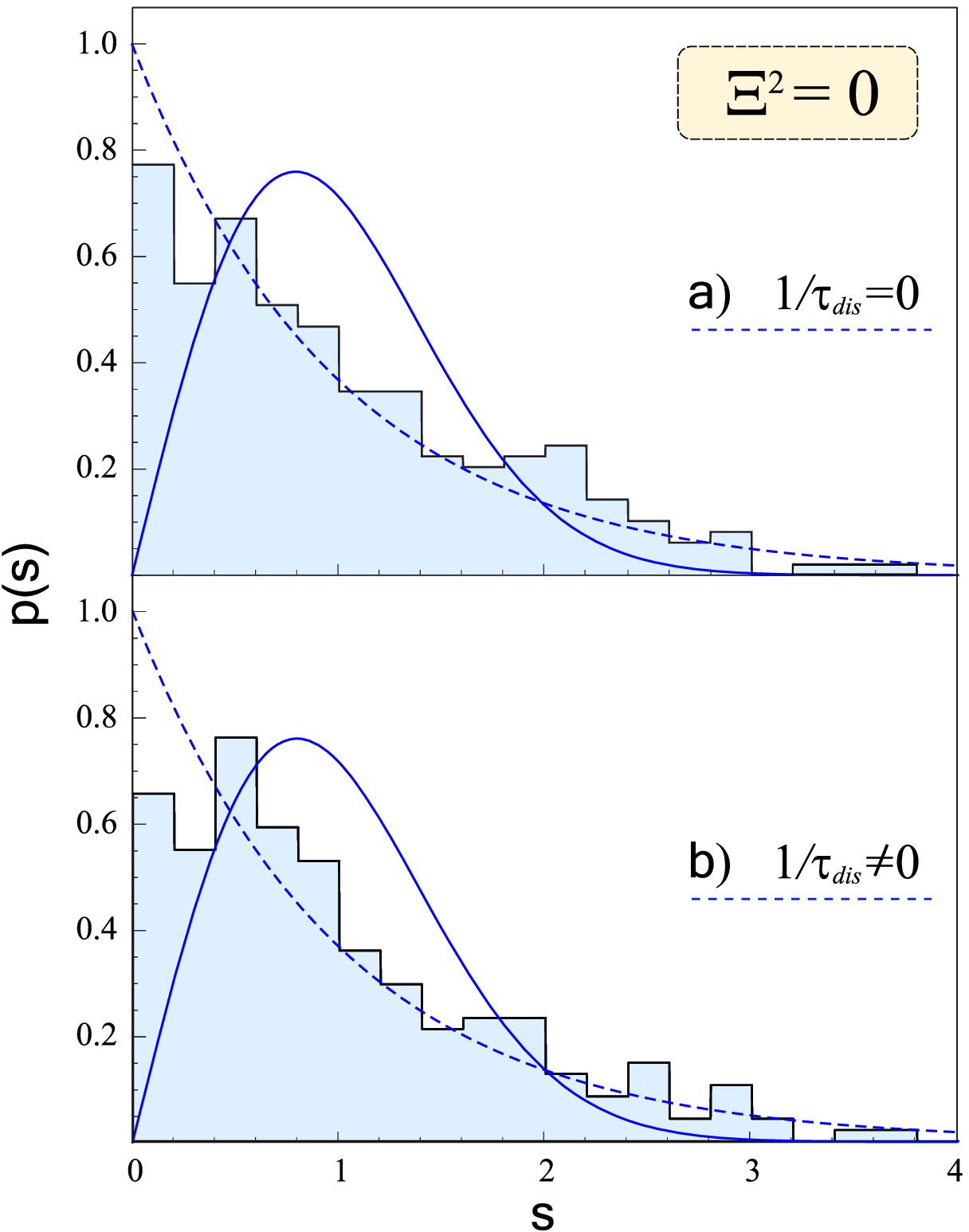}
  \hspace{.5cm}%
  \includegraphics{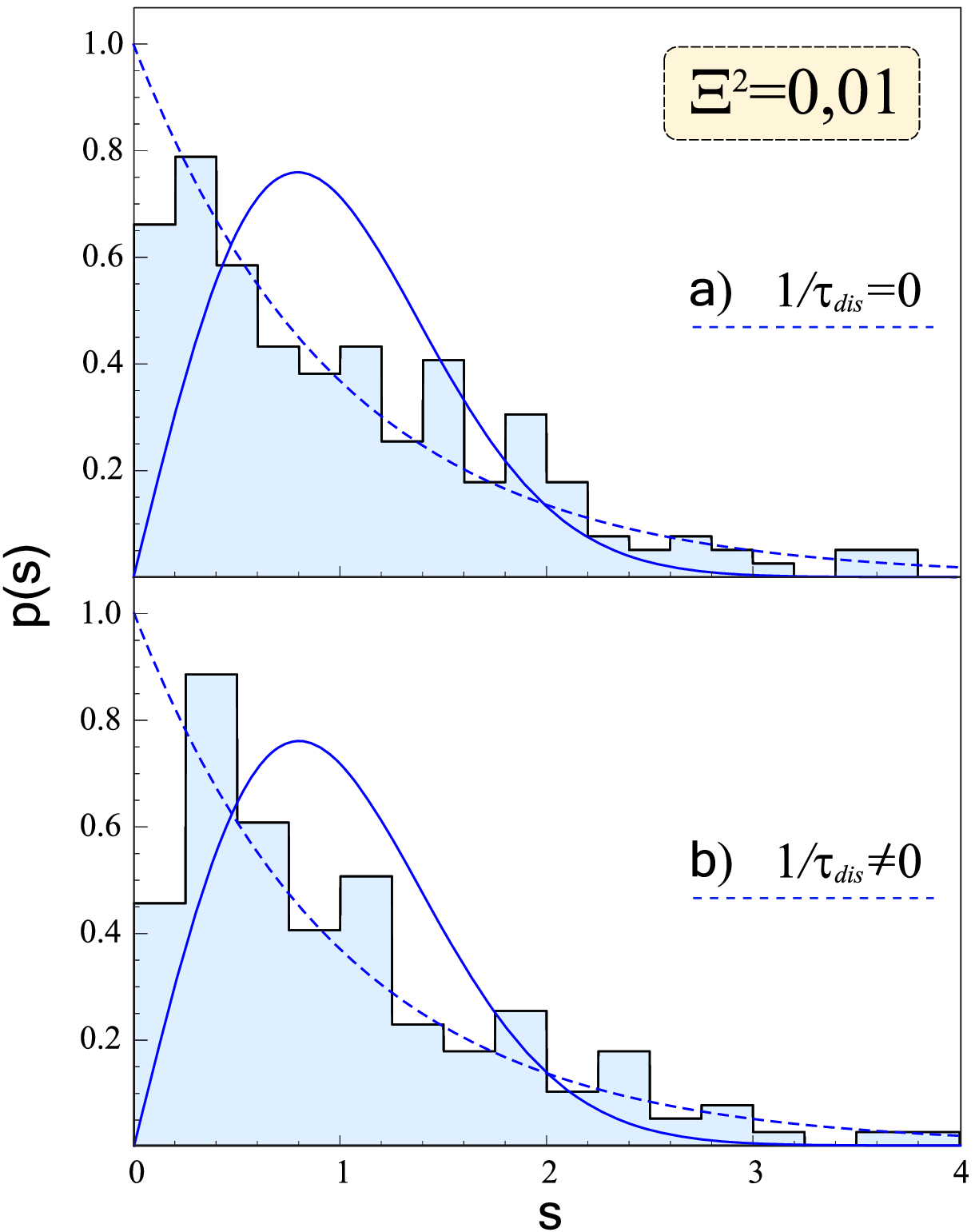}
  \hspace{.5cm}%
  \includegraphics{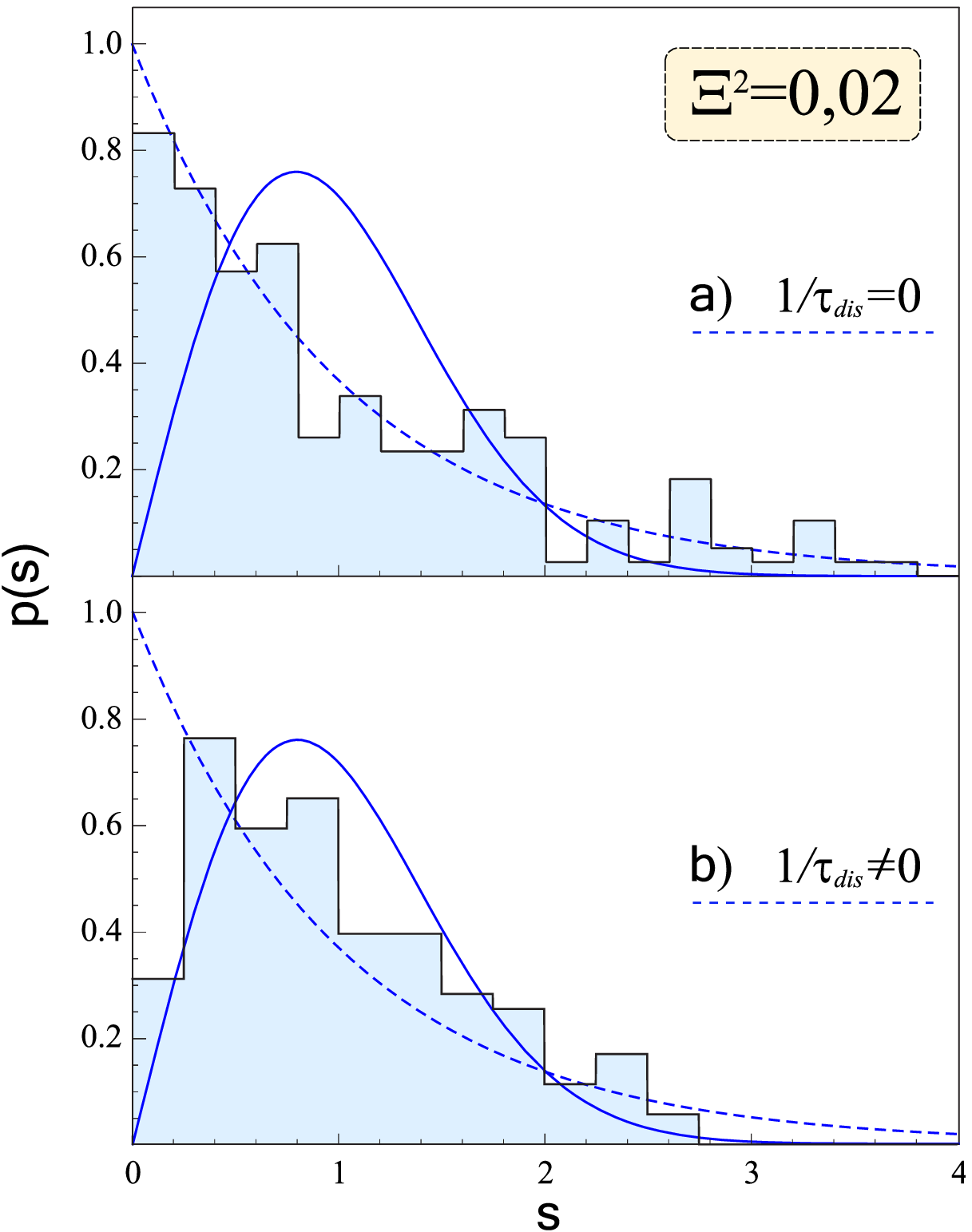}
  \hspace{.5cm}%
  \includegraphics{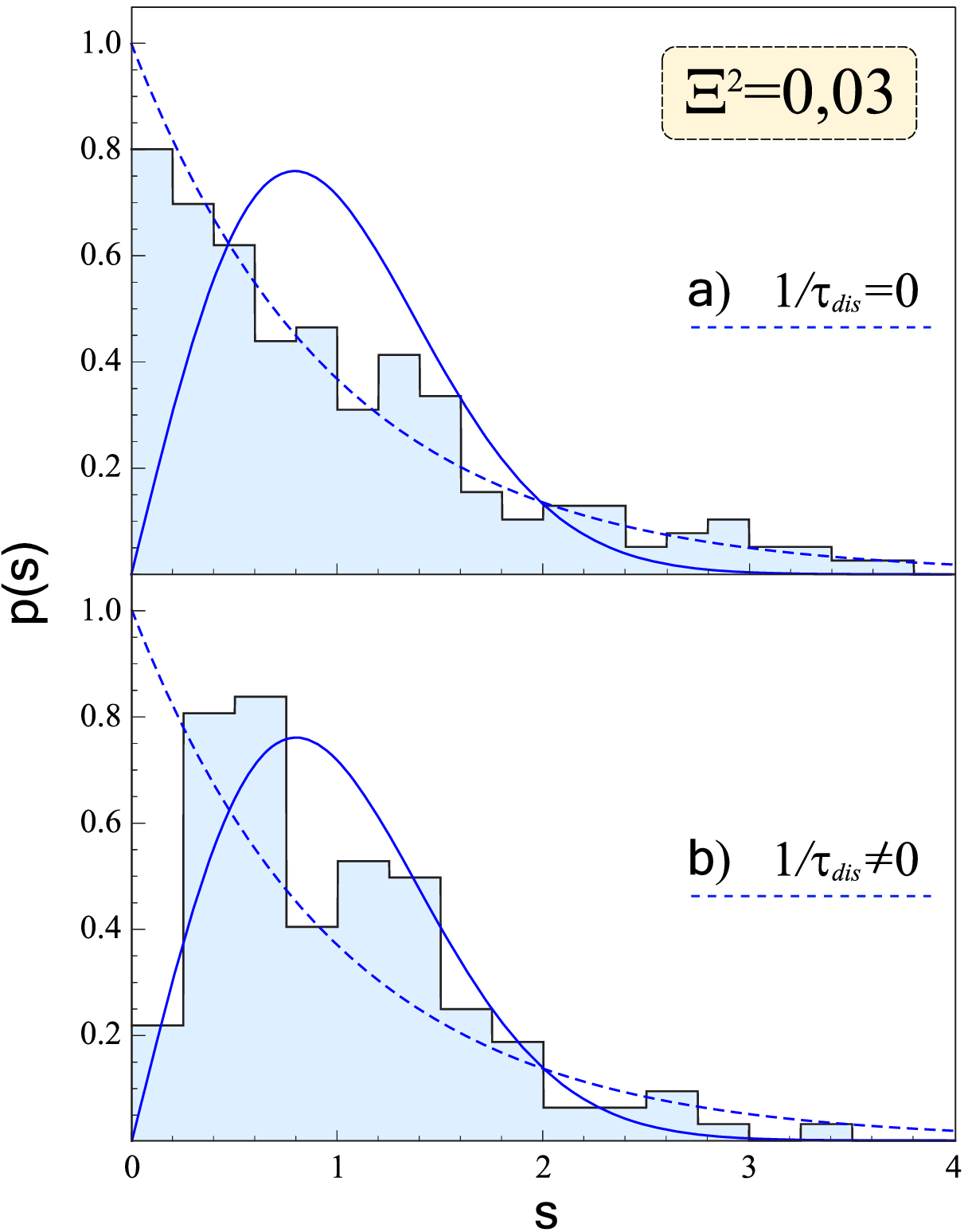}  }
  \caption{NLS distribution for different values of slope parameter $\Xi^2$ in the small-slope domain. Histograms are obtained from Eqs.~\eqref{DeltaKappa+1/tau}. In each diagram, Wigner and Poisson distributions (solid and dashed lines, respectively) are shown for reference purposes. \hfill
 \label{fig2}}
\end{figure*}

Figure~\ref{fig2}, shows NLS histograms for inhomogeneity-free resonator side-walls ($\Xi=0$) and for some instances where parameter $\Xi$ is of finite but small value, as compared to unity. It is evident that if the dissipation loss is not taken into account (upper plots in Fig.~\ref{fig2})
then the level spacings are distributed close to Poisson law, however rough the resonator may be. Considering the dissipation and, consequently, the resonance line finite width leads to partial ``Wignerization'' of the NLS distribution, specifically, to the appearance of a significant dip of the function $p(s)$ in the region of $s\approx 0$. This is due to the fact that, when counting the number of resonances in a given frequency window, only resonance frequencies spaced at distances larger than the width of the corresponding lines should be taken into account. All other resonances, which are positioned very closely with a given one, in our calculations were considered as merging into a single resonance peak.

\begin{figure*}[h]
  \captionstyle{centerlast}
  \centering
  \scalebox{.65}[.65]{\includegraphics{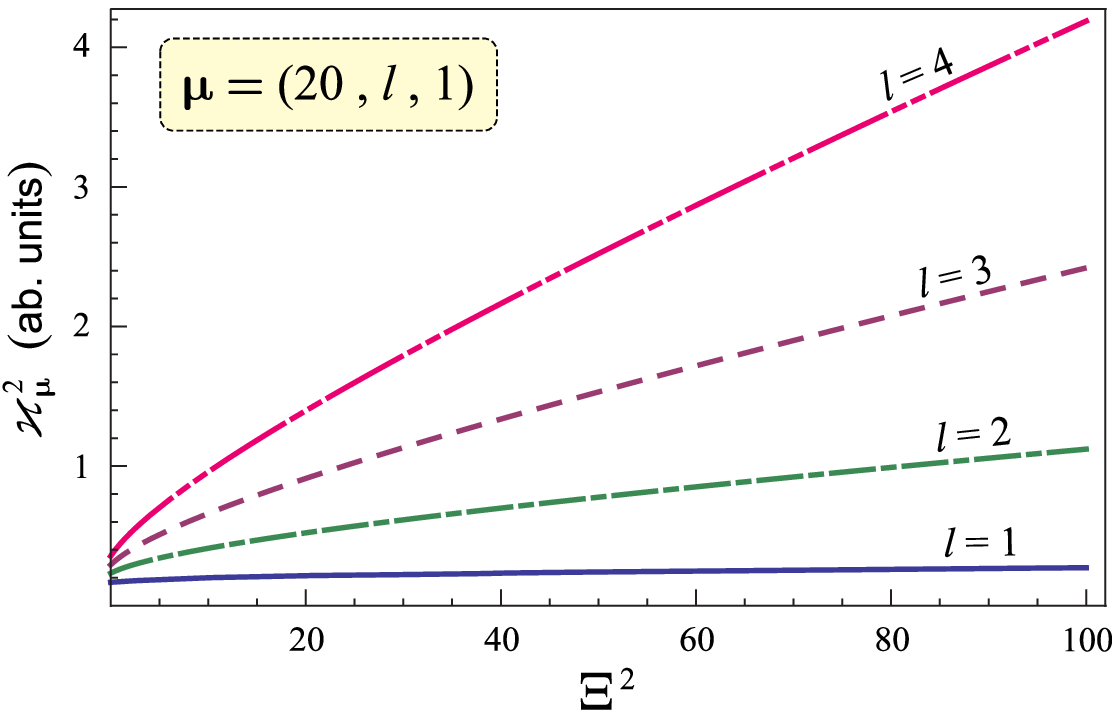}
  \hspace{1.5cm}%
  \includegraphics{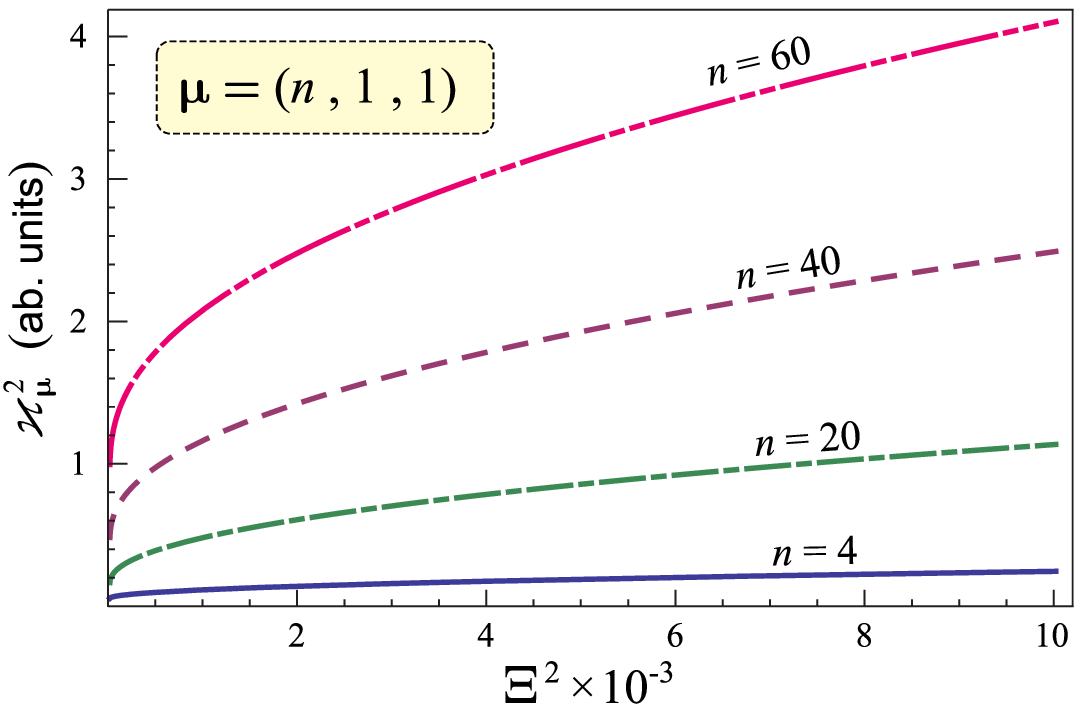} }
  \caption{The dependence of composite mode energy \eqref{mode_en} on the degree of sharpness of side boundary asperities. \hfill
 \label{fig3}}
\end{figure*}

On gradual sharpening the asperities, specifically, in the domain where $\Xi\gg 1$, the resonance frequency distribution may well be described by formula \eqref{mode_en} only. The broadening related to roughness-induced mode scattering (the analog of Eq.~\eqref{1/tau} for $\Xi\ll 1$) monotonically decreases with growing parameter $\Xi$, and in the limit $\Xi\to\infty$ the width of all resonance lines is again determined by solely the dissipation loss. Figure~\ref{fig3} shows the dependence of several composite mode energies on the asperity sharpness. With increasing parameter $\Xi$ all the energies monotonically increase. The intervals between neighboring levels tend to increase as well.

This particular fact allows one to neglect (under the condition of strong scattering, $\Xi\gg 1$) not only the scattering-induced decoherence but also the initial dissipative broadening, which was chosen to be small from the outset so as to resolve the bulk of the resonance lines. As can be seen from the graphs in Fig.~\ref{fig3}, the composite modes ``float up'' in energy as the asperities get sharpened. In each prescribed frequency interval the number of resonances decreases with growing $\Xi$, going to zero as this parameter tends to infinity.

\begin{figure*}[h]
  \captionstyle{centerlast}
  \centering
  \scalebox{.45}[.45]{\includegraphics{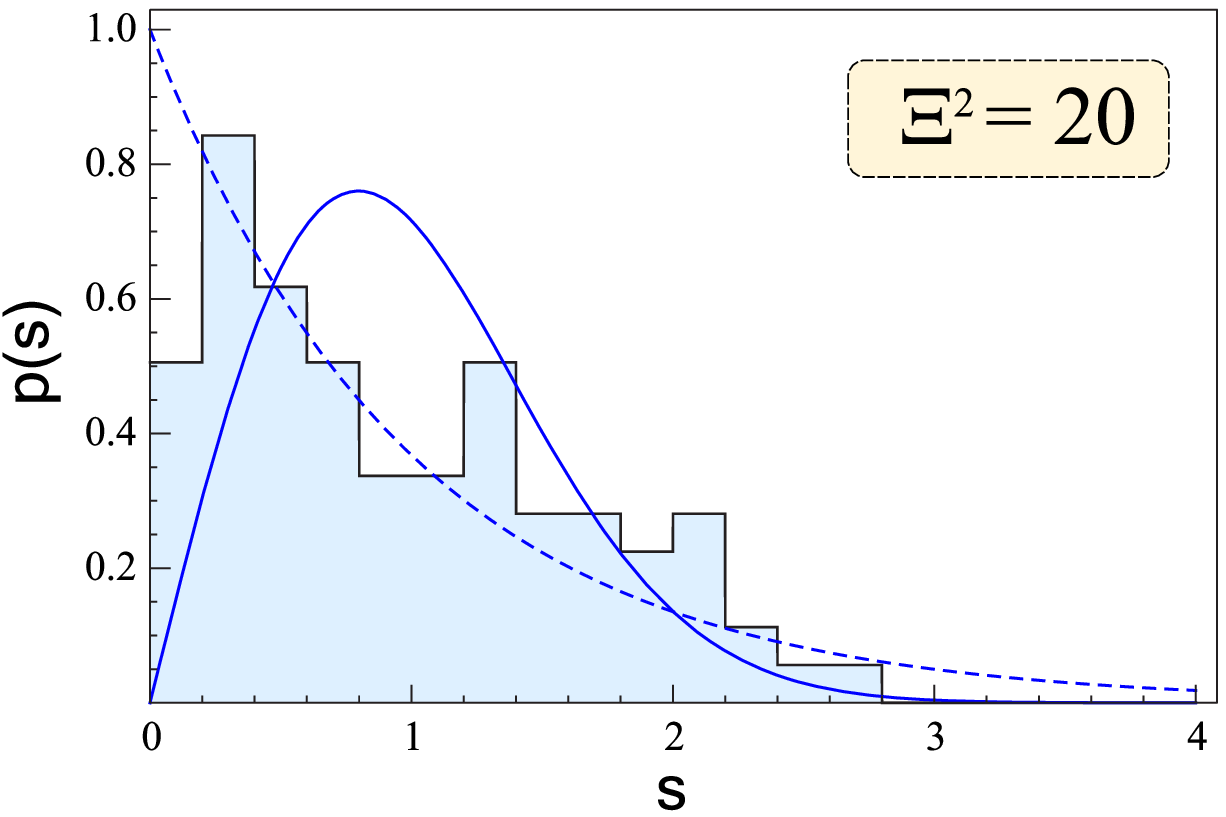}
  \hspace{1cm}%
  \includegraphics{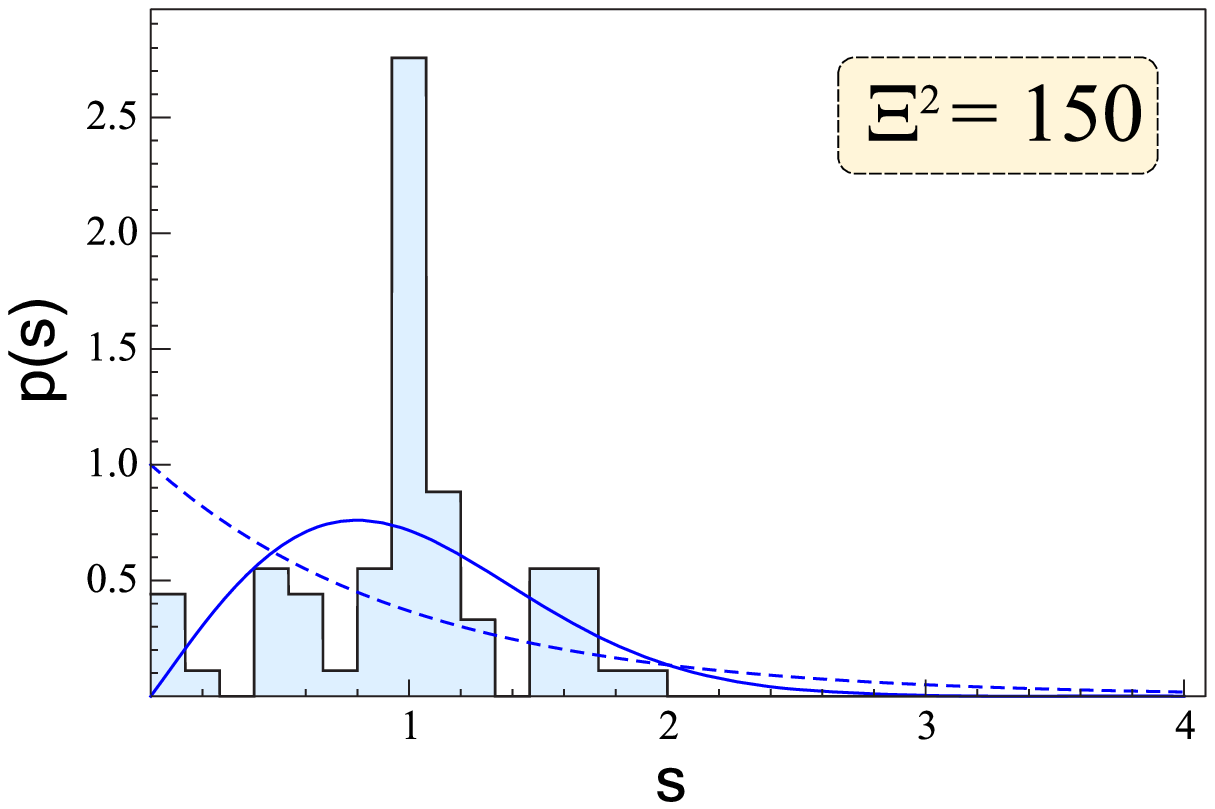}
  \hspace{1cm}%
  \includegraphics{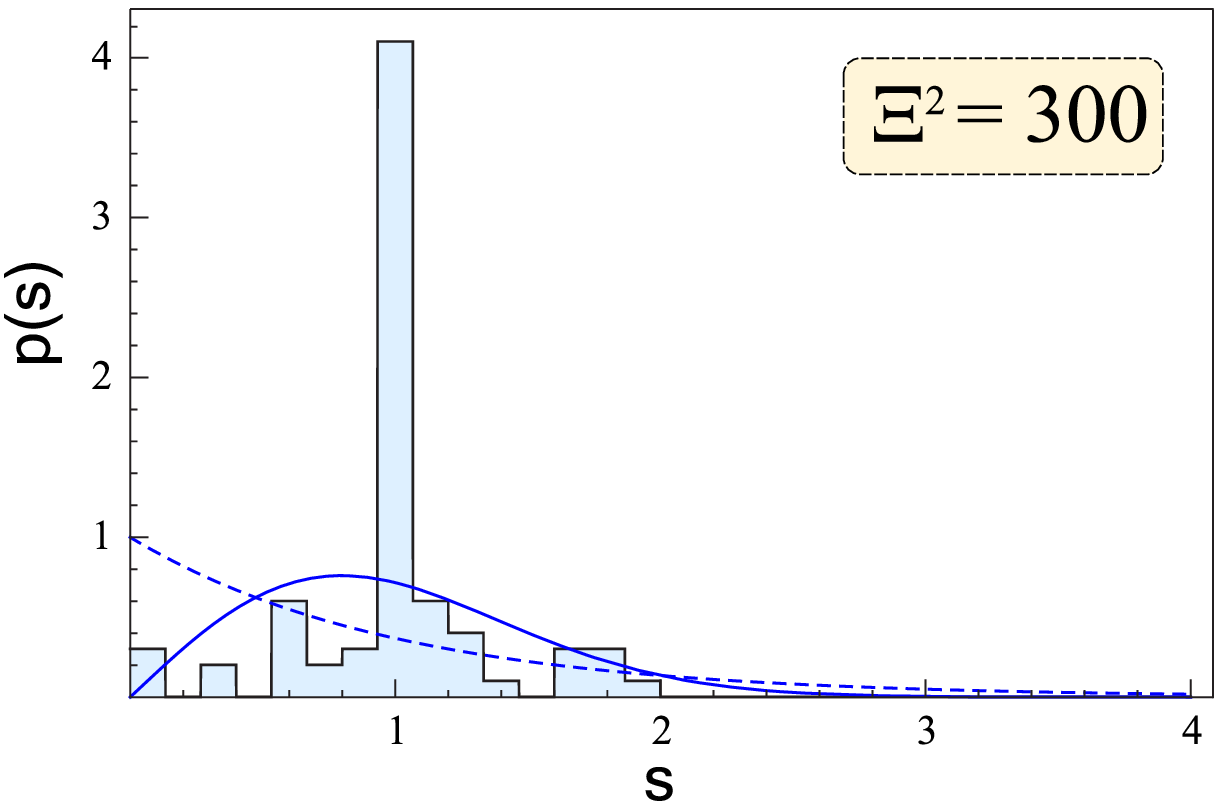} }
  \caption{NLS distributions for large values of the slope parameter $\Xi^2$. Gradual level clustering in the vicinity of $s=1$ is clearly seen in the case where mean sharpness of the asperities measures extremely large values. \hfill
 \label{fig4}}
\end{figure*}

In Fig.~\ref{fig4}, NLS histograms for very sharp boundary asperities are given, when composite mode scattering is
particularly strong.
One can see that on sharpening the boundary irregularities the interfrequency interval distribution, being of Poisson nature at small $\Xi$ and in the absence of dissipation, transforms gradually to the bell-shaped curve with a maximum placed at $s=1$. With $\Xi\to\infty$ this distribution, allowing for its normalization, approaches the $\delta$ function, namely, $\delta(s-1)$. Such a form is characteristic for NLS distributions of \emph{a fortiori} integrable systems, in particular, of a set of harmonic oscillators with commensurable frequencies. Note that this particular system can always be specified with a single governing parameter; i.\,e., it can be thought of to be, in a way, one-dimensional.

Such a behavior of NLS distribution for a randomly rough resonator, which is at first glance unusual, can be explained
in the following way. As an example, consider a resonator with no dissipation and follow the transformation of $p(s)$ with an increase in the gradient parameter~$\Xi$. At the early stage, in the region of small (as compared to unity) values of this parameter, the distribution retains Poissonian form  (upper plots in Fig.~\ref{fig2}), which may be accepted as a signature of small interlevel correlations. In this case, the set of energy levels of weakly rough two-dimensional wave billiard (we consider the case of a single $z$ mode, $k=1$) in any reasonably large frequency interval is governed by \emph{two} parameters (two mode-index components), as one can see from Eq.~\eqref{mode_en}.

With growing the parameter $\Xi$, the NLS distribution gradually transforms from the distribution of the Poisson type,
with the maximum at $s=0$, to the bell-shaped distribution whose maximum moves up to the point $s=1$ (see Fig.~\ref{fig4}). In doing so, in view of the above discussed ``float-up'' effect, the system of resonance levels in any prespecified frequency interval approaches ever more the one-parameter set, which is characteristic of regular physical systems. Indeed, the effective (gradient-renormalized) azimuth indices entering eigenenergy \eqref{mode_en} for oscillations with different $l$'s simultaneously go to zero when $\Xi$ unlimitedly increases, and resonances tend to
become differentiated (quantized) by the radial index only. If azimuth indices of all oscillations were really strictly
the same we would get the $\delta$-shaped distribution, $p(s)=\delta(s-1)$, instead of actual peaked distributions with finite widths.

The finite width of the actually obtained peaked distributions is accounted for by the peculiar fact that at arbitrary large  but finite values of the parameter $\Xi$ one can always indicate frequency intervals (maybe, of quite large frequency values) where the bulk of resonance levels having the same azimuth index $n$ and distinguished by radial index only is supplemented with a certain number of levels (the lesser is the number the larger is $\Xi$) with different azimuth indices. In the classical phase space such admixing of ``foreign'' levels results in the deformation of energy contours out of the phase plane, which normally is associated with the onset of chaos. Meanwhile, in the wave mechanics we deal with in the present work the resonance levels, even though intermixed, in the absence of dissipation remain perfectly stable (i.\,e., of zero width). This is the main reason why we cannot speak about the presence of \emph{quantum} (or \emph{wave}) \emph{chaos} in the rough-side resonator unless it is dissipative.

The analogous situation was already discussed in the paper by Berry and Tabor (see Ref.~\cite{bib:BerryTabor77}), where the authors, using the definitely integrable system of harmonic oscillators as an example, have put forward and substantiated the concept of \emph{level clustering} in the vicinity of either $s=0$ or $s=1$. Obviously, the distribution of single-oscillator energy levels, which are entirely correlated, obeys the law $p(s)=\delta(s-1)$. It was exactly the presence of other oscillators with incommensurable frequencies (the condition preventing the set of energy levels from being one-parametric) that has led the authors of Ref.~\cite{bib:BerryTabor77} to the NLS distributions similar in form to those depicted in Fig.~\ref{fig4} for two largest values of the gradient parameter, namely, for $\Xi^2=150$ and $\Xi^2=300$.

\section{Discussion}

In summary, we have suggested a new calculation method for finding the spectra of resonators with nonhomogeneous
boundaries. The method is based on direct solution of wave equations rather than on the system symmetry properties, and
it is tried out for a cylindrical cavity resonator with a randomly rough side boundary (randomly rough two-dimensional wave billiard). Although in the context of the classical-billiard-system theory this resonator is obviously nonintegrable, being considered from a position of wave/quantum theory its governing equation is shown to be separable, even if the system boundary is of quite complicated form. The condition mandatory for the corresponding multidimensional wave equation to be separable is the violation of either $\mathcal{P}$ or $\mathcal{T}$ symmetry of the wave operator. For the inhomogeneities of random nature $\mathcal{P}$ symmetry is manifestly broken. This made it possible for us to study both dissipative and nondissipative rough cavity resonators within the unified approach.

Using exact mode separation, which is carried out with our method at an arbitrary level of the disorder at resonator boundaries, it is shown that the quality factors of resonance lines can become finite only if there is dissipation and/or radiation loss in the system. In the absence of dissipation, despite the boundary complexity, the
oscillation spectrum of randomly rough resonator appears to be regular and possesses no signs of chaos.

We also have found that the intensity of intermode scattering, and thus the degree of entanglement of resonator modes, is predominantly determined by the average sharpness of surface asperities rather than by their mean height. In the
absence of dissipation and under conditions of weak oscillation scattering produced by the resonator wall roughness (i.\,e., in the case of smooth boundary asperities) the interfrequency interval distribution in the resonator spectrum is close to Poissonian distribution. The resonance lines, being strictly discrete when there is no dissipation, with the presence of the latter become manifestly widened. The lines positioned tightly on the frequency axis begin to merge with one another and cannot thus be considered as independent ones. The distribution of intervals between neighboring resonances broadened concurrently by dissipation and roughness-induced scattering acquires a significant Wigner component. The numerical estimations we have carried out on the basis of our formulas for nonsharp asperities, Eqs.~\eqref{DeltaKappa+1/tau}, are in good qualitative agreement with our recent experimental results reported in Ref.~\cite{bib:GanapTarasov11}.

Within the suggested approach we were also able to consider, besides the case of smoothly rough resonators, the
resonators whose boundary inhomogeneities are arbitrarily sharp. The sharpening of the asperities results in the increase in energies of the effective (composite) oscillation modes, which gradually leave upward of any frequency interval chosen for spectral measurements. The spectrum of the quasioptical resonator, being initially dense within any sufficiently large frequency range, in view of the mode ``floating-up'' effect that arises due to boundary inhomogeneities is substantially rarefied, thereby approaching monofrequency, or even zero-frequency regime if the asperities become sufficiently sharp.

The results we have obtained in this study are pertinent to billiard-type wave systems, which are ballistic from the
viewpoint of classical dynamics. The conclusions we arrived at differ drastically from those obtained previously for resonators of similar cylindric form, though randomly inhomogeneous in the bulk infill rather than on the surface. The wave transport in disordered resonators with continuously varying bulk permittivity is always of diffusion nature~\cite{bib:GanErTar06-1,bib:GanErTar06-2,bib:GanErTar07},\footnote{The classic analogs of such resonators are nonconservative dynamic systems, where energy is not conserved in time.} irrespective of whether the dissipation is taken into account or not. For this reason, the state of chaos in their spectrum, which is determined, according to the Chirikov criterium, by the resonance line width~\cite{bib:ZaslChir72,bib:Chirikov79}, is always finite. Hence, the
conclusion results that the spectra of wave and quantum resonance systems with continuous inhomogeneities in their
bulk are always to some extent chaotic, whereas the analogous systems of truly billiard configuration reveal chaotic properties only provided that there exist some dissipation channels.

\begin{acknowledgments}
This work is partially supported by the Ministry of Education and Science of Ukraine under the project ``Fundamental Problems of Nanostructured Systems, Nanomaterials, and Nanotechnologies'', Project No.~0107U003985.
\end{acknowledgments}

\vspace{\baselineskip}

\appendix

\section{Evaluation of the mode-mixing operator norm}
\label{Intermix-norm}

When estimating the norm of the mode-mixing operator $\hat{\mathsf R}$ through Eq.~\eqref{R-norm_estim}, for the trial Green's function $G_{\bm{\nu}}^{(V)}$ we apply expression \eqref{G_trial} with the intramode potential $\mathcal{V}_{\bm{\nu}}$ put equal to zero based upon arguments presented in Sec.~\ref{Intramode_scat-estim}.

As to the intermode scattering, to estimate the contribution to the norm \eqref{R-norm_estim} of the height potential $\hat{V}^{(h)}$ one has to evaluate the correlator $\Av{{\mathcal U}^{(h)*}_{\bm{\nu'}\bm{\mu}}{\mathcal U}^{(h)}_{\bm{\nu}\bm{\mu}}}$. Substituting the potential $\hat{V}^{(h)}$ into Eq.~\eqref{U_mode} in the asymptotic form \eqref{HeigtPot-appr} and performing the averaging we obtain
\begin{widetext}
\begin{equation}\label{U_numu(h)}
  \Av{{\mathcal U}^{(h)*}_{\bm{\nu'}\bm{\mu}}
  {\mathcal U}^{(h)}_{\bm{\nu}\bm{\mu}}}=
  4\left(\frac{\sigma}{R}\right)^2
  \left[k^2-\left(\frac{\pi q_{\bm{\mu}}}{H}\right)^2\right]^2
  \widetilde{W}(n_{\bm{\nu}}-n_{\bm{\mu}})
  \delta_{\bm{\nu}\bm{\nu'}}
  \delta_{q_{\bm{\nu}}q_{\bm{\mu}}}
  \delta_{l_{\bm{\nu}}l_{\bm{\mu}}}\ .
\end{equation}
Then, upon inserting expression Eq.~\eqref{U_numu(h)} into Eq.~\eqref{R-norm_estim} and taking account of the normalization $\sum_n\widetilde{W}(n)=1$ we arrive at the following estimate for the height-potential contribution to the intermixing operator $\hat{\mathsf R}$,
\begin{equation}\label{R(h)-norm_estim}
  \Av{\|\hat{\mathsf R}^{(h)}\|^2}\sim
  \left(\frac{\sigma}{R}\right)^2\ll 1\ .
\end{equation}

The correlator $\Av{{\mathcal U}^{(s1)*}_{\bm{\nu'}\bm{\mu}}{\mathcal U}^{(s1)}_{\bm{\nu}\bm{\mu}}}$, which arises in estimating the norm of the gradient-potential operator $\hat{\mathsf R}^{(s1)}$, after averaging can be recast as follows:
\begin{equation}\label{<Us1Us1>}
  \Av{{\mathcal U}^{(s1)*}_{\bm{\nu'}\bm{\mu}}{\mathcal U}^{(s1)}_{\bm{\nu}\bm{\mu}}}=\left(\frac{\sigma}{R}\right)^2
  \delta_{q_{\bm{\nu}'}q_{\bm{\nu}}}\delta_{n_{\bm{\nu}'}n_{\bm{\nu}}}
  \delta_{q_{\bm{\nu}}q_{\bm{\mu}}}\left(n^2_{\bm{\nu}}-n^2_{\bm{\mu}}\right)^2
  \widetilde{W}(n_{\bm{\nu}}-n_{\bm{\mu}})
  \int_0^Rdr'\bra{r';l_{\bm{\nu}'}}\frac{\partial}{\partial r'}
  \ket{r';l_{\bm{\mu}}}^*
  \int_0^Rdr\bra{r;l_{\bm{\nu}}}\frac{\partial}{\partial r}
  \ket{r;l_{\bm{\mu}}}\ .
\end{equation}
It is easy to verify, subject to the particular form of eigenfunction \eqref{Eigenfunc_2}, that the product of integrals in Eq.~\eqref{<Us1Us1>} is appropriately estimated by the following expression:
\begin{equation}\label{r_r'_int-estim}
  \int_0^Rdr'\bra{r';l_{\bm{\nu}'}}\frac{\partial}{\partial r'}
  \ket{r';l_{\bm{\mu}}}^*
  \int_0^Rdr\bra{r;l_{\bm{\nu}}}\frac{\partial}{\partial r}
  \ket{r;l_{\bm{\mu}}}\sim
  \frac{1}{R^4}\cdot\frac{\gamma_{l_{\bm{\mu}}}^{(|\widetilde{n}_{\bm{\mu}}|)}}%
  {\gamma_{l_{\bm{\mu}}}^{(|\widetilde{n}_{\bm{\mu}}|)}+
  \gamma_{l_{\bm{\nu}}}^{(|\widetilde{n}_{\bm{\nu}}|)}}\cdot
  \frac{\gamma_{l_{\bm{\mu}}}^{(|\widetilde{n}_{\bm{\mu}}|)}}%
  {\gamma_{l_{\bm{\mu}}}^{(|\widetilde{n}_{\bm{\mu}}|)}+
  \gamma_{l_{\bm{\nu}'}}^{(|\widetilde{n}_{\bm{\nu}}|)}}\ .
\end{equation}
For estimation purposes we set $\widetilde{n}_{\bm{\mu}}\sim kR$ in Eq.~\eqref{r_r'_int-estim} and then, based upon the relationship $\sqrt{|\widetilde{n}|(|\widetilde{n}|+2)}<\gamma_1^{(|\widetilde{n}|)}<  \sqrt{2|\widetilde{n}|(|\widetilde{n}|+1)}$ known from the Bessel function theory \cite{bib:Watson49}, we use the inequality $\forall\gamma_{l}^{(|\widetilde{n}_{\bm{\mu}}|)}\gtrsim
kR$. Substituting the asymptotic expression
\begin{equation}\label{gamma_l^n-asymp}
  \gamma_{l}^{(|\widetilde{n}|)}\approx
  \left(l+\frac{|\widetilde{n}|}{2}-\frac{3}{2}\right)\pi\hspace{1cm}
  \big(l\gg |\widetilde{n}|\big)
\end{equation}
into Eq.~\eqref{r_r'_int-estim} instead of the exact roots $\gamma_{l}^{(|\widetilde{n}|)}$, one can make
sure that replacing the product of integrals in Eq.~\eqref{<Us1Us1>} with the product \eqref{r_r'_int-estim} not only enables one to make a correct estimate of each of the terms in the sum \eqref{R-norm_estim} but also to carry out
twofold summation over radial mode indices. The estimation formula for the operator $\hat{\mathsf R}^{(s1)}$ norm is then reduced to
\begin{align}\label{R(s1)-estim}
  \Av{\|\hat{\mathsf R}^{(s1)}\|^2}\sim \max_{\bm{\mu}}\frac{1}{(kR)^2}\left(\frac{\sigma}{R}\right)^2 
  \sum_{n_{\bm{\nu}}}(n_{\bm{\mu}}^2-n_{\bm{\nu}}^2)^2
  \widetilde{W}(n_{\bm{\mu}}-n_{\bm{\nu}})\ ,
\end{align}
whereupon we arrive at the outcome
\begin{equation}\label{R^(s1)-fin_est}
   \Av{\|\hat{\mathsf R}^{(s1)}\|^2}\sim
   \left(\frac{\sigma}{s_c}\right)^2\sim\Xi^2\ .
\end{equation}

To estimate the norm of the operator $\hat{\mathsf R}^{(s2)}$ originating from the square-gradient potential \eqref{SlPot-2}one has to evaluate the following correlator:
\begin{align}
   \Av{{\mathcal U}^{(s2)*}_{\bm{\nu'}\bm{\mu}}
   {\mathcal U}^{(s2)}_{\bm{\nu}\bm{\mu}}}= &
   \delta_{q_{\bm{\nu}'}q_{\bm{\nu}}}\delta_{q_{\bm{\nu}}q_{\bm{\mu}}}
   \frac{4}{R^4}B_{l_{\bm{\nu}}}^{(\widetilde{n}_{\bm{\nu}})}
   B_{l_{\bm{\nu}'}}^{(\widetilde{n}_{\bm{\nu}'})}
   \left[B_{l_{\bm{\mu}}}^{(\widetilde{n}_{\bm{\mu}})}\right]^2\notag\\
 & \times  \int_0^1tdt\,
   J_{|\widetilde{n}_{\bm{\nu}'}|}
   \left(\gamma_{l_{\bm{\nu}'}}^{(|\widetilde{n}_{\bm{\nu}'}|)}t\right)
   \left[\left(\gamma_{l_{\bm{\mu}}}^{(|\widetilde{n}_{\bm{\mu}}|)}\right)^2-
   \frac{\widetilde{n}_{\bm{\mu}}^2}{t^2}\right]
   J_{|\widetilde{n}_{\bm{\mu}}|}
   \left(\gamma_{l_{\bm{\mu}}}^{(|\widetilde{n}_{\bm{\mu}}|)}t\right)\notag\\
 & \times  \int_0^1t'dt'\,
   J_{|\widetilde{n}_{\bm{\nu}}|}
   \left(\gamma_{l_{\bm{\nu}}}^{(|\widetilde{n}_{\bm{\nu}}|)}t'\right)
   \left[\left(\gamma_{l_{\bm{\mu}}}^{(|\widetilde{n}_{\bm{\mu}}|)}\right)^2-
   \frac{\widetilde{n}_{\bm{\mu}}^2}{{t'}^2}\right]
   J_{|\widetilde{n}_{\bm{\mu}}|}
   \left(\gamma_{l_{\bm{\mu}}}^{(|\widetilde{n}_{\bm{\mu}}|)}t'\right)\notag\\
 & \times \oint\frac{d\varphi}{2\pi}\mathrm{e}^{i(\widetilde{n}_{\bm{\nu'}}-
   \widetilde{n}_{\bm{\mu}})\varphi}
   \oint\frac{d\varphi'}{2\pi}\mathrm{e}^{-i(\widetilde{n}_{\bm{\nu}}-
   \widetilde{n}_{\bm{\mu}})\varphi'}
   \left\{\frac{1}{R^4}\AV{\big[\xi'(\varphi)\xi'(\varphi')\big]^2}-\Xi^4\right\}
   \ .
 \label{<U_numuU_nu'mu>^s2}
\end{align}
The integrals over $\varphi$ and $\varphi'$ can be easily calculated with the use of the Gaussian model for the random process $\xi(\varphi)$, and as a consequence we have
\begin{align}
   \oint\oint\frac{d\varphi d\varphi'}{(2\pi)^2}\ldots
   = & \frac{2}{R^4}\oint\oint\frac{d\varphi d\varphi'}{(2\pi)^2}
   \exp\big[-i(\widetilde{n}_{\bm{\nu}'}-
   \widetilde{n}_{\bm{\mu}})\varphi+
   i(\widetilde{n}_{\bm{\nu}}-
   \widetilde{n}_{\bm{\mu}})\varphi'\big]
   W^2(\varphi-\varphi')\notag\\
   = & 2\left(\frac{\sigma}{R}\right)^4\delta_{n_{\bm{\nu}'}n_{\bm{\nu}}}
   \sum_{k=-\infty}^{\infty}\widetilde{W}(k)
   \widetilde{W}(k+\widetilde{n}_{\bm{\mu}}-\widetilde{n}_{\bm{\nu}})\ .
 \label{Ints_phiphi'}
\end{align}
\end{widetext}
In addition, if one takes the function $W(\varphi)$ in the form of a Gaussian exponential, then $\widetilde{W}(k)=(\varphi_c/\sqrt{2\pi})\exp(-k^2\varphi_c^2/2)$, and the double integral over azimuth angles in Eq.~\eqref{Ints_phiphi'} are estimated by the simple formula
\begin{equation}\label{Ints(phiphi')-estim}
   \oint\oint\frac{d\varphi d\varphi'}{(2\pi)^2}\ldots\sim
   \varphi_c\left(\frac{\sigma}{R}\right)^4\exp\left[-\frac{(\widetilde{n}_{\bm{\mu}}-
   \widetilde{n}_{\bm{\nu}})^2\varphi_c^2}{4}\right]\ .
\end{equation}

The integrals  in Eq.~\eqref{<U_numuU_nu'mu>^s2}, which contain the Bessel functions, can be estimated by substituting the asymptotic expressions valid for large arguments for those functions. The sums over radial mode indices are then easily calculated, and one obtains the following estimate for the operator $\hat{\mathsf R}^{(s2)}$ square norm:
\begin{equation}\label{R(s2)-estim}
   \Av{\|\hat{\mathsf R}^{(s2)}\|^2}\sim
   \left(\frac{\sigma}{R}\right)^4\varphi_c
   \max_{\bm{\mu}}\sum_{\widetilde{n}_{\bm{\nu}}}
   \exp\left[-\frac{(\widetilde{n}_{\bm{\mu}}-
   \widetilde{n}_{\bm{\nu}})^2\varphi_c^2}{4}\right]\ .
\end{equation}
In so far as the angle correlation parameter is small in magnitude, $\varphi_c\ll 1$, one can approximately replace the sum in Eq.~\eqref{R(s2)-estim} with the integral, thereby obtaining
\begin{equation}\label{R^(s2)-fin_est}
   \Av{\|\hat{\mathsf R}^{(s2)}\|^2}\sim
   \left(\frac{\sigma}{R}\right)^4\ll 1\ .
\end{equation}
From this estimation it follows that the quadratic in the $\xi'(\varphi)$ potential in Eq.~\eqref{GreenEqMain}, in contrast to the linear potential $\hat{V}^{(s1)}$, is adequately taken into account through the substitution into all
of the formulas of the mean value of this potential.


\end{document}